\documentclass[reprint,pra,twocolumn,showpacs,superscriptaddress,aps,longbibliography]{revtex4-2}

\usepackage{graphicx}
\usepackage{epstopdf}

\usepackage[T1]{fontenc}
\usepackage[applemac]{inputenc}
\usepackage{lmodern}
\usepackage[english]{babel}

\usepackage{ae}
\usepackage{siunitx}

\usepackage{amsmath,amssymb,natbib,bm}
\usepackage{amsfonts}
\usepackage{psfrag}
\usepackage{subfigure}
\usepackage{amsthm}
\usepackage{algorithm}
\usepackage{algpseudocode}
\usepackage{listings}
\usepackage{multirow}
\usepackage{physics}
\usepackage{tikz}

\usepackage[colorlinks]{hyperref}
\hypersetup{%
        plainpages=true,
        breaklinks=true,
        hypertexnames=false,
        pageanchor=true,
        colorlinks=true,
        linkcolor={blue},
        citecolor={blue},
        urlcolor={blue},
        anchorcolor={black}
      }

\usepackage{mleftright} 

\newcommand{\figref}[1]{\mbox{Fig.~\ref{#1}}}

\newcommand{\secref}[1]{\mbox{Sec.~\ref{#1}}}

\newcommand{\appref}[1]{\mbox{Appendix~\ref{#1}}}
\renewcommand{\eqref}[1]{\mbox{Eq.~(\ref{#1})}}

\newcommand{\figpanel}[2]{Fig.~\hyperref[#1]{\ref*{#1}(#2)}}
\newcommand{\figpanels}[3]{Fig.~\hyperref[#1]{\ref*{#1}(#2)--(#3)}}
\newcommand{\figpanelNoPrefix}[2]{\hyperref[#1]{\ref*{#1}(#2)}}

\begin{document}

\title{Heuristically optimizing, synthesizing, and prioritizing \\ measurement settings for quantum state tomography}

\author{Sumukh~S.~Moudghalya}
\affiliation{Department of Microtechnology and Nanoscience, Chalmers University of Technology, 41296 Gothenburg, Sweden}

\author{Anton Frisk Kockum}
\affiliation{Department of Microtechnology and Nanoscience, Chalmers University of Technology, 41296 Gothenburg, Sweden}

\author{Akshay Gaikwad}
\email{akshayga@chalmers.se}
\affiliation{Department of Microtechnology and Nanoscience, Chalmers University of Technology, 41296 Gothenburg, Sweden}


\begin{abstract}

A key task in many quantum-computing applications, e.g., quantum simulation and quantum state tomography (QST), is to partition an arbitrary set of operators into mutually commuting subsets and efficiently measure these subsets. However, brute-force approaches to this task quickly become intractable as the number and dimensionality of operators grow. In this article, we reformulate operator partitioning as a graph coloring (GC) problem and develop an efficient computational framework to solve it, balancing accuracy and computational efficiency. Our framework enables leveraging a range of GC algorithms, including degree of saturation, recursive largest first, integer linear programming, graph neural networks, and spectral clustering. We first benchmark these algorithms for operator partitioning. Then, we demonstrate their utility in both optimizing QST experiments, where determining non-overlapping data acquisition settings for QST is a major challenge, and prioritizing among these settings, i.e., selecting the experiments that provide the most information. We further show how to perform these experiments by directly synthesizing Clifford circuits for joint measurement of commuting Pauli operators in multi-qubit systems, using a binary symplectic representation and symplectic Gaussian elimination. The framework is validated across multi-qubit (up to five qubits), multi-qutrit (up to three qutrits), and hybrid qubit-qutrit systems. Our results show that heuristic GC methods substantially reduce the number of required measurement settings for QST and enable priority-based scheduling that maximizes the information gain per experiment. The optimization converges within minutes on a student-grade laptop (8-core CPU, 16 GB RAM, no dedicated GPU), providing speedups of several orders of magnitude over brute-force methods already for these relatively small quantum systems. This demonstrates the potential of GC heuristics as a scalable and practical tool for characterization of noisy intermediate-scale quantum devices. We have made the Python implementation of our GC framework to optimize and schedule QST experiments publicly available at \url{https://github.com/ssm8015/QST_GT.git}.

\end{abstract}

\date{\today}

\maketitle


\section{Introduction}

Quantum state tomography (QST) is the procedure to reconstruct the density matrix of a quantum system from measurement data~\cite{james-pra-2001, liu-prb-2005, lvovsky-rmp-2009, cramer-prl-2013}. It provides a complete description of the underlying quantum state. Operationally, QST is typically performed in two stages: (i) \emph{data acquisition}, in which expectation values of a prescribed set of observables are measured, and (ii) \emph{data post-processing}, in which a reconstruction algorithm maps the measurement data to a physical density matrix. Although conceptually straightforward, QST rapidly becomes impractical as the system size grows, due to an exponential increase in both the number of required experiments and the computational cost of reconstruction~\cite{cotler-prl-2020, riofrio-natcom-2017, li-pra-2017}.

To mitigate this prohibitive complexity of QST, a broad range of tomography protocols has been developed~\cite{Gebhart2023, Hashim2025, Blume-Kohout2025, mle-pra-2007, imle, hsu-prl-2024, wang-prr-2024, quek-npj-2021, Ahmed-cgan-2021, lohani-mlst-2020, gaikwad-pra-2024, david-prl-2010, Steffens_2017, gaikwad-arxiv-2025, tangyou-prl-2025}. Many of these advanced protocols focus on the post-processing step, e.g., improving estimators, incorporating physical constraints, or accelerating computation for final reconstruction. However, the equally important question of \emph{how to choose, and efficiently implement, measurement settings for data acquisition}, has received less attention~\cite{flammia-prl-20111, huang-natphys-2020, gaikwad-qip-2022, anushya-prl-2024}. In this article, our goal is therefore to construct, in an efficient way, an experimentally feasible set of global measurement settings that is tomographically complete and uses as few experiments as possible, for systems residing in arbitrary-dimensional Hilbert space.

Existing measurement-optimization strategies often target specific tasks, e.g., estimating selected density-matrix elements or restricted properties, rather than performing full tomography~\cite{bonet-prx-2020, gaikwad-epjd-2023, kim-natcom-2017, luca-prl-2018, lundeen-prl-2016, zhang-prl-2019, paz-pra-2013, Bolduc-natcom-2016, feng-pra-2021, mahesh-arxiv-2025, patel-arxiv-2025}. When extended to full QST, these approaches can become inefficient---and sometimes even more demanding than standard schemes---in terms of both the number of experimental configurations and practical implementation. Although minimal measurement requirements for \emph{pure}-state tomography have been identified~\cite{goyeneche-prl-2015, feng-arxiv-2025}, they do not directly apply to general mixed states. 

\begin{figure*}
    \centering
    \includegraphics[width=0.8\linewidth]{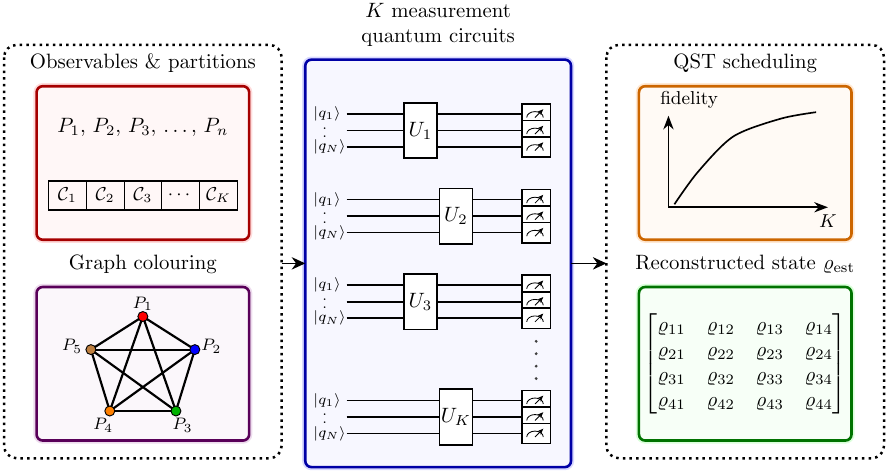}
    \caption{Schematic workflow of measurement scheduling for quantum state tomography using graph coloring. First, an informationally complete set of quantum observables $\{ P_1, P_2, P_3, \ldots, P_n\}$ is partitioned, using heuristic GC algorithms, into $K$ commuting subsets $\mathcal{C}_1, \mathcal{C}_2, \ldots, \mathcal{C}_K$, which correspond to measurement settings. For each such subset (color class), corresponding quantum circuits with unitaries $\{U_1, U_2, U_3, \ldots, U_K\}$ are compiled. The resulting measurement data are acquired and processed using a scheduling strategy, illustrated as the fidelity improving with the number of executed settings $K$, to reconstruct an estimate $\varrho^{\rm est}$ of the density matrix describing the system state.}
    \label{fig:flowchart}
\end{figure*}

Other lines of work based on overlapping tomography~\cite{cotler-prl-2020} and adaptive QST methods~\cite{quek-npj-2021, lange-quantum-2023} address the challenge of optimizing measurement strategies by reducing redundancy in experimental data. Recently, multi-qubit QST has been demonstrated using only a small set of Pauli measurements~\cite{chai-prapplied-2023}. Another conceptually appealing option is measurement in mutually unbiased bases (MUBs), which maximizes information gained per setting and has strong theoretical and experimental support~\cite{ivono-math-1981, wootters-ap-1989, robin-jmp-2004, adamson-prl-2010, wang-pra-2024}. However, global MUB constructions are restricted to prime or prime-power Hilbert-space dimensions; they may be unavailable in composite dimensions or in arbitrary subspaces. For example, Ref.~\cite{rao-physscr-2019} constructs optimized measurement operators for certain multi-level spin systems, but the procedure relies on the existence of MUBs for the relevant dimension.

More generally, any set of linearly independent Hermitian operators spanning the operator space is informationally complete and can be used for QST. The practical bottleneck is \emph{how to measure such operator sets efficiently}. The multi-qubit Pauli operator set is a canonical example, but measurement strategies that are experimentally convenient and that extend beyond local measurements remain limited, especially when the relevant state lives in a reduced or otherwise arbitrary-dimensional multi-qubit subspace. In such scenarios, measuring the full Pauli operator set is unnecessary---one should instead re-optimize the experimental configurations for the reduced operator set. 

In other words, even though the recipe for efficiently partitioning the complete set of Pauli operators is reported and very well studied~\cite{reggio-pra-2024}, the problem of efficiently partitioning an arbitrary subset of Pauli operators---or, more generally, arbitrary operators---into commuting subsets is still a daunting task, even for multi-qubit systems of moderate size. The number of operators grows quickly with system size, and the ways to partition them, in turn, grows combinatorially---brute-force approaches are thus infeasible. Similar challenges arise beyond qubits: for multi-qutrit, general multi-qudit, and hybrid composite systems. The task we seek to perform can thus be formulated as efficiently finding the optimal experimental settings required to estimate multiple selective elements of the density matrix of a quantum state residing in a target subspace---or to perform full QST---in a Hilbert space of arbitrary dimension.

Here, we tackle this task using graph-coloring (GC) heuristics to both optimize and prioritize among measurement configurations for full or partial QST of arbitrary quantum states in Hilbert spaces of any dimension. The workflow for this approach to QST is schematically shown in \figref{fig:flowchart}. As illustrated on the left in that figure, we recast the problem of minimizing the number of measurement settings as a partitioning problem over mutually commuting subsets of operators. This partitioning task can be naturally translated into a graph-coloring problem: vertices represent operators, edges connect commuting (or non-commuting) pairs, and each color class corresponds to a commuting operator set that can be measured simultaneously in a common eigenbasis. As a result, all operators within one color class can be estimated within a single experimental configuration, reducing the number of experiments and accelerating the overall tomography procedure.

Efficiently partitioning large operator sets (beyond multi-qubit Pauli operators) is computationally challenging and remains an active area of research. Prior studies on Pauli partitioning~\cite{gokhale-arxiv1907.13623, reggio-pra-2024, temme-quantum-2020, jena-pra-2022, jena-arxiv-2019, sarkar-mathsci-2021, kurita-jphyschem-2023} have analyzed structural properties and demonstrated the value of near-optimal partitions in applications beyond QST, such as the variational quantum eigensolver and Hamiltonian simulation. Related graph-theoretic ideas, such as clique-cover formulations, have been used to optimize overlapping-tomography experiments for reconstructing $k$-body marginals of an $N$-qubit state~\cite{kiara-arxiv-2024}, with recent experimental demonstrations~\cite{wei-prappl-2025}.

We show that the GC heuristics provide a general framework applicable to \emph{any} operator set. Within this framework, we study and benchmark several algorithms---including degree of saturation (DSATUR), recursive largest first (RLF), integer linear programming (ILP), graph neural networks (GNNs), and spectral clustering (SC)---to balance solution quality and runtime. We evaluate the performance of these algorithms in terms of (i) the number of partitions (measurement settings) they produce and (ii) their runtime and computational scaling. Furthermore, by using a binary symplectic representation of Pauli operators and symplectic Gaussian elimination, we directly synthesize change-of-basis quantum circuits consisting of only elementary Clifford gates (Hadamard $H$, phase gate $S$, and CNOT) for joint measurement of the resulting partitions to perform full QST (see the middle of \figref{fig:flowchart}).

From numerical simulations, we find that the GC-based approach can substantially reduce the number of required quantum circuits (i.e., minimize the number of measurement settings). Furthermore, it also enables priority-based scheduling that maximizes information gain per setting and avoids data overlap (see the right part of \figref{fig:flowchart}). We validate the framework on multi-qubit (up to five qubits), multi-qutrit (up to three qutrits), and hybrid qubit-qutrit systems. In all cases, the optimization converges within minutes on a standard laptop, yielding orders-of-magnitude speedups over brute-force partitioning already for these comparatively small quantum systems. 

We have made the Python implementation of the GC algorithms developed in this work publicly available at \url{https://github.com/ssm8015/QST_GT.git}. The code operates in three main stages.
(i) First, it takes as input the target operator set and partitions it into relatively few subsets of mutually commuting operators. 
(ii) Next, it generates quantum circuits or unitary operations designed to measure each of these subsets. 
(iii) Finally, it processes the measurement data obtained from these circuits to estimate the desired density-matrix elements or to perform partial or full QST.

This article is structured as follows. In \secref{sec:qst}, we first review the standard procedure for QST to set the stage for our work. Then, in \secref{sec:gc-qst}, we introduce the proposed GC algorithms for partitioning a set of operators into subsets of mutually commuting operators. As the final part of the methods section, we provide in \secref{sec:scheduling} a description of the measurement-scheduling and -implementation strategy for QST using our techniques. In \secref{sec:partition-results}, we present results for operator partitioning using the GC methods, with corresponding QST performance obtained using the proposed priority-based measurement scheduling framework reported in \secref{sec:qst-results}. Finally, \secref{sec:conclusion} concludes the paper by summarizing our findings and outlining directions for future work.


\section{Methods}
\label{methods}

To put the partitioning of sets of operators in context, we here first provide an overview of the standard workflow for QST, encompassing both the data-acquisition and data-processing stages. With this background in place, we then show how the selection of QST experiments (partitioning arbitrary sets of operators) maps to a GC problem, and discuss in detail how several common GC algorithms can be applied to this problem. Finally, we show how insights from solving the GC problem can be used to systematically prioritize among QST experiments, and how these experiments translate into quantum circuits.


\subsection{Quantum state tomography}
\label{sec:qst}

The objective of QST is to reconstruct an unknown density matrix $\varrho \in \mathbb{C}^{d \times d}$, which describes the quantum state of a system in a $d$-dimensional Hilbert space. Such a density matrix is generally expressed as
\begin{equation} \label{eq:rho-def}
\varrho = \sum_{i,j=0}^{d-1} \varrho_{ij} \ketbra{i}{j},\hspace{0.15cm}{\rm s.t.} \hspace{0.15cm} \varrho  = \varrho^\dag, \hspace{0.15cm} {\rm Tr}(\varrho) = 1, \hspace{0.15cm} \& \hspace{0.15cm} \varrho \geq 0 ,
\end{equation}
where $\varrho_{ij}$ represents the element of $\varrho$ in the $i$th row and $j$th column. Here, $\{\ketbra{i}{j}\}$ forms an operator basis set of cardinality $d^2$, reflecting the number of independent real parameters required to uniquely characterize $\varrho$, with $\{ \ket{i} \}$ being the underlying computational basis states.


\subsubsection{Data acquisition}
\label{sec:DataAcquisition}

To reconstruct an unknown $\varrho$, QST requires measurement data $\{ \mathcal{B}_i \}$ and an appropriate post-processing algorithm $\mathcal{J}$ that takes $\{ \mathcal{B}_i \}$ as input and compute $\varrho$ as output, i.e., $\mathcal{J}: \{ \mathcal{B}_i \} \rightarrow \varrho$. The measurement data $\{ \mathcal{B}_i \}$ is often expressed as expectation values of a set of observables $\mathcal{P}= \{ P_1, P_2, \cdot \cdot\cdot, P_k \}$, such that $\mathcal{B}_i = \text{Tr}(P_i \varrho)$. In the most general scenario, $\mathcal{P}$ can be a set of positive operator valued measures (POVMs), a set of Hermitian operators, or simply a set of projection operators~\cite{miranowicz-pra-2014, robin-jmp-2004, sacchi-job-2004}. 

In most current experimental setups, the standard procedure to estimate the expectation value of a given observable $P_i$ involves applying an appropriate change-of-basis operation ${U}_i$, followed by a measurement in the computational basis set $\Pi = \{ \ket{0}\bra{0}, \ket{1}\bra{1}, \cdot \cdot \cdot, \ket{d}\bra{d} \}$. This procedure, commonly referred to as measurement in the $P_i$ eigenbasis, can be expressed as
\begin{equation}\label{eq:data}
\mathcal{B}_i = \mathrm{Tr}(P_i \varrho) = \sum_j \lambda_j^i \mathrm{Tr}[ U_i \varrho U_i^\dag \Pi_j ],
\end{equation}
with ${U}_i^\dag \Pi_j {U}_i = \ketbra{\lambda_j^i}{\lambda_j^i}$, where $\ket{\lambda_j^i}$ is the $j$th normalized eigenvector of $P_i$ and $\lambda_j^i$ is the corresponding eigenvalue. 

A set of unitary operations $\mathcal{U} = \{ {U}_1, {U}_2, \cdot \cdot \cdot, {U}_k \}$ is said to be \textit{tomographically} or \textit{informationally complete} (IC) if the set $\{ {U}_i^{\dagger} \Pi_j {U}_i \:\vert\: {U}_i \in \mathcal{U} \:\&\: \Pi_j \in \Pi \}$ spans the entire operator space. This condition can be verified by evaluating the rank of the sensing matrix $\mathcal{S}$ associated with the measurements. This sensing matrix has dimension $k d \times d^2$ with $k = \vert \mathcal{U}\vert$ and can be constructed analytically:
\begin{equation}
\mathcal{S}_{mn} = \text{Tr}(M_m K_n) ,
\end{equation} 
where $ M = \{ U_m^\dag \Pi_k U_m \:\vert\: U_m \in \mathcal{U}, \Pi_k \in \Pi \}$ and  $K = \{\ketbra{i}{j} \:\vert\: i,j = [0, d-1] \}$.
If $\rank(\mathcal{S}) = d^2$, then the set $\mathcal{U}$ together with $\Pi$ is IC. 

To illustrate these concepts, consider an $N$-qubit system. For this system, one of the possible IC sets of unitary operations, based solely on local transformations, is
\begin{equation} 
\label{nqubit_ic}
\mathcal{U} = \{ I, R_x, R_y \}^{\otimes N},
\end{equation}
where $I$ denotes the $2 \times 2$ identity matrix, and $R_x$ and $R_y$ are single-qubit rotation operators for $\pi/2$ rotations about the $x$ and $y$ axes, respectively. This set satisfies the IC condition since $\rank(\mathcal{S}) = 4^N$. In other words, the unitaries in \eqref{nqubit_ic} enable the estimation of all $N$-qubit Pauli operators $\mathcal{P} = \{ I, X, Y, Z \}^{\otimes N}$, where $X$, $Y$, and $Z$ are the standard single-qubit Pauli matrices. The cardinality of this IC set is $3^N$, increasing exponentially with the number of qubits, implying that $3^N$ experimental settings are required to obtain an IC data set using these observables. This choice of observables and unitary operations is currently the standard one for QST experiments in most quantum-computing platforms. However, this set is not optimal when global operations are permitted. 

In this article, we address the problem of identifying and constructing optimal IC sets $\mathcal{U}$ composed of global measurement operations for arbitrary-dimensional quantum systems. Doing so involves partitioning the desired operator set into mutually commuting subsets (see the left part of \figref{fig:flowchart}), where each subset $i$ corresponds to a common eigenbasis and therefore can be measured simultaneously using a single unitary transformation $U_i$. For the particular case of multi-qubit systems, the problem of partitioning the complete set of Pauli operators into mutually commuting subsets has been extensively studied from a mathematical standpoint~\cite{reggio-pra-2024, temme-quantum-2020}, with emphasis on the algebraic structure and properties of these subsets~\cite{jena-pra-2022, jena-arxiv-2019, reggio-pra-2024, sarkar-mathsci-2021}. However, these algebraic methods are specific to the structure of Pauli operators and do not naturally generalize to arbitrary operator sets. Moreover, they may become inefficient when applied to reduced or otherwise arbitrary subsets of Pauli operators encountered in practical scenarios.


\subsubsection{Data processing}

After the measurement data set $\{ \mathcal{B}_i \}$ has been acquired according to \eqref{eq:data}, the data is processed to complete the QST. In the data-processing step, an appropriate algorithm is used to reconstruct a physically valid density matrix from the data, ensuring that all three validity constraints in \eqref{eq:rho-def} are satisfied. Mathematically, this reconstruction can be formulated as the following constrained convex optimization problem~\cite{gaikwad-qip-2021}:
\begin{subequations}
\begin{align} 
\min_{ \{\varrho^{\text{est}} \} } \quad & \sum_{i,j} \mleft[ \mathcal{B}_i - \mathrm{Tr}(P_i \varrho^{\text{est}}) \mright]^2 \label{eq:ls-cco} \\
\text{s.t.} \quad & \varrho^{\text{est}} \geq 0, \\
                        & \mathrm{Tr}(\varrho^{\text{est}}) = 1.
\end{align}
\end{subequations}
This type of optimization problem can be efficiently solved for small- to moderate-dimensional systems using established convex-optimization frameworks such as CVX or YALMIP, which provide convenient interfaces and rely on robust built-in solvers~\cite{cvx-solvers}.

To handle somewhat larger systems, one can employ more efficient optimization algorithms, e.g., stochastic gradient descent with physically valid density-matrix parameterizations, as demonstrated in Ref.~\cite{gaikwad-arxiv-2025}. However, since the focus of this article is on developing efficient measurement strategies for the data-acquisition stage, we do not give more details on these data-processing algorithms here. Instead, we refer readers interested in advanced reconstruction techniques to Ref.~\cite{gaikwad-arxiv-2025} for a more comprehensive discussion.


\subsection{Graph-coloring algorithms for quantum state tomography}
\label{sec:gc-qst}

Having seen in \secref{sec:DataAcquisition} that data acquisition for QST can be improved by cleverly choosing which measurements to make, we now show how the problem of making this choice maps to the problem of GC. We then present several heuristic algorithms for GC, and show how they can be applied to measurement selection in QST. This completes the methods required for the left part of our QST workflow in \figref{fig:flowchart}.


\subsubsection{The selection of measurements in quantum state tomography as a graph-coloring problem}
\label{qst_gc}

Here, we reformulate the task of dividing a set of arbitrary operators $\mathcal{P}$ into mutually commuting subsets---an essential step for efficient QST---as a GC problem. First of all, we note that the problem of GC is to color the elements of a graph such that no two connected elements share the same color, using as few colors as possible. The minimum number of colors solving this problem is called the chromatic number. In our mapping to QST here, each color class corresponds to a subset of mutually commuting operators, and reaching the chromatic number would correspond to finding the minimum number of measurements required for the QST.

To carry out this mapping, we construct an undirected graph $\mathcal{G}_{\mathcal{P}} = (\mathcal{V}, \mathcal{E})$, in which each vertex $v_i \in \mathcal{V}$ uniquely represents an operator $P_i \in \mathcal{P}$, such that $\vert \mathcal{V}\vert =\mathcal{K}(\mathcal{P})$, where $\mathcal{K}(\mathcal{P})$ is the cardinality of $\mathcal{P}$. An edge $e_{ij} \in \mathcal{E}$ connects vertices $v_i$ and $v_j$ if and only if the corresponding operators $P_i$ and $P_j$ do not commute, i.e., $[P_i, P_j] \neq 0$. With this construction, the adjacency matrix $\mathcal{A}$ is thus given by
\begin{equation} 
    \mathcal{A}_{ij}(\mathcal{G_P}) = 
       \begin{cases}
         1, & \text{if } [P_i,P_j]\neq0,\\
         0, & \text{if } [P_i,P_j]=0.
       \end{cases}
       \label{adj_mat}
\end{equation}

The adjacency matrix $\mathcal{A}(\mathcal{G_P})$ defines the structure of the graph $\mathcal{G_P}$ and serves as the input to any GC algorithm. These algorithms assign colors to the vertices such that no two adjacent vertices (i.e., those connected by an edge) share the same color, while aiming to minimize the total number of colors used, denoted by $\chi(\mathcal{G_P})$. Since the decision version of the GC problem is NP-complete and the optimization version is NP-hard, there is no known polynomial-time algorithm that can solve it optimally in the general case~\cite{garijo-robust-2024,hecht-approx-random-2023}. The quality and feasibility of a solution depend on several factors, including whether an exact or heuristic approach is desired, the size and structure of the graph, and the available computational resources. In this article, since the number of operators that need to be partitioned grows rapidly with system size, we focus on heuristically minimizing $\chi(\mathcal{G_P})$ to achieve a balance between solution optimality and computational efficiency.


\subsubsection{Degree of saturation and recursive largest first algorithms for graph coloring}

We now give an overview of such heuristic GC algorithms, starting with the widely used \textit{degree of saturation} (DSATUR)~\cite{Daniel-ACM-79, barry-operational_research-10, yekezare-orl-2024} and \textit{recursive largest first} (RLF)~\cite{leighton-graph_coloring_heuristic-79, david-cliques_coloring-96, carter-comp_eact_heuristic-06}. These two algorithms have polynomial time complexities of $\mathcal{O}(n^2)$ and $\mathcal{O}(n^3)$, respectively, where $n$ denotes the number of vertices. They have been extensively applied in various domains, including combinatorial scheduling, frequency assignment in wireless networks, and register allocation in compilers.

The DSATUR algorithm determines an effective ordering of vertices for color assignment based on the saturation degree, which is defined as the number of distinct colors assigned to adjacent vertices. In contrast, RLF builds color classes recursively, prioritizing vertices with the highest degree (i.e., the largest number of neighbors). We provide more formal pseudocode descriptions of DSATUR and RLF in Algorithm~\ref{dsatur} and Algorithm~\ref{rlf}, respectively, in \appref{app:Pseudocode}.

\begin{figure}
    \centering
    \includegraphics[width=\linewidth]{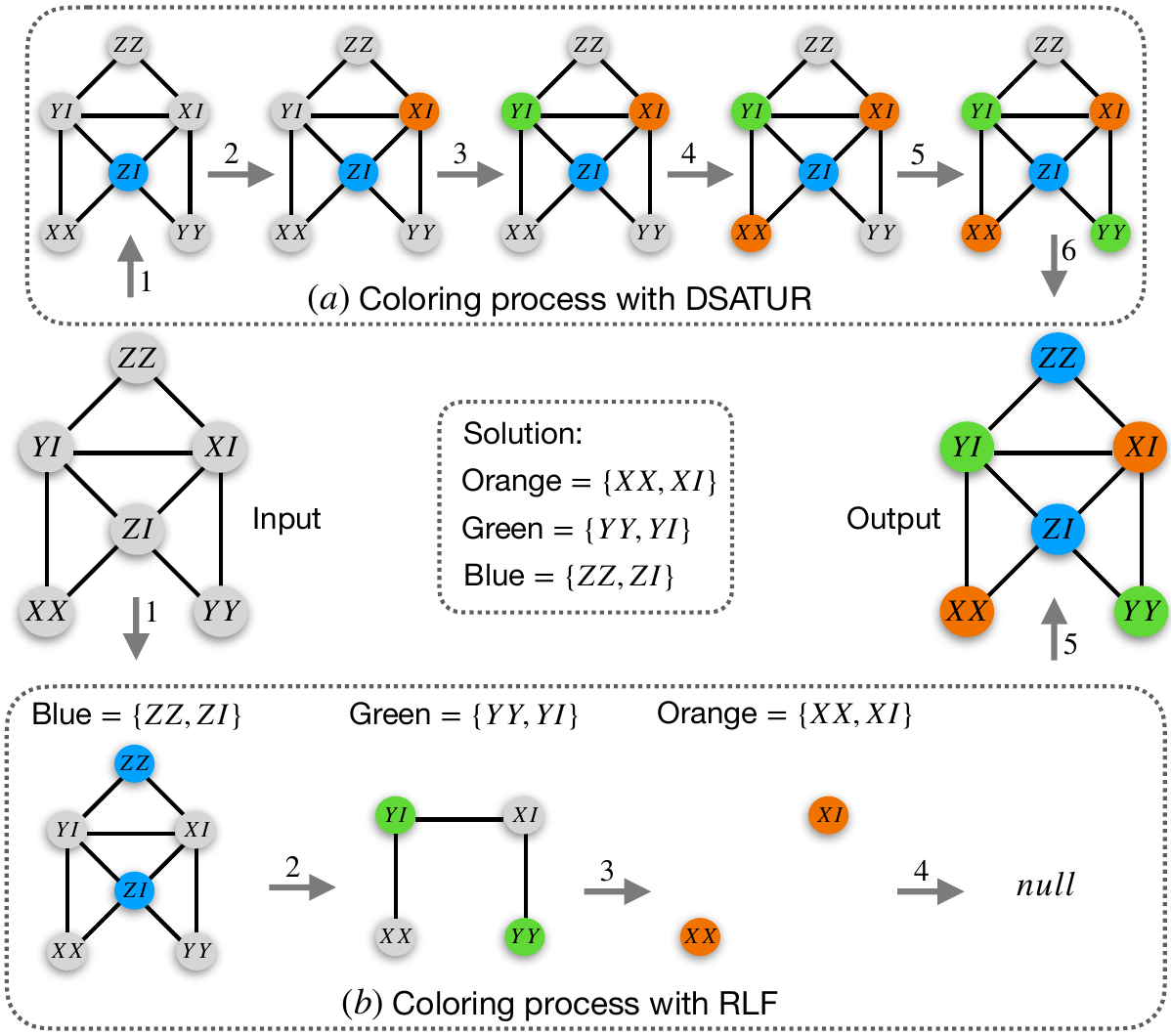}
    \caption{Pictorial representation of the graph coloring process using
    (a) DSATUR (upper panel) and
    (b) RLF (bottom panel) applied to the two-qubit Pauli-operator subset $\mathcal{P} = \{ XI, YI, ZI, XX, YY, ZZ \}$ as an example.}
    \label{fig:dsatur_rlf_coloring}
\end{figure}

To provide a clearer understanding of these approaches to solving GC and how they apply to operator partitioning, \figref{fig:dsatur_rlf_coloring} illustrates how the DSATUR [\figpanel{fig:dsatur_rlf_coloring}{a}] and RLF [\figpanel{fig:dsatur_rlf_coloring}{b}] algorithms are applied to an example with a subset of two-qubit Pauli operators, $\mathcal{P} = \{ XI, YI, ZI, XX, YY, ZZ \}$. In the graph, each vertex represents a Pauli operator, and an edge between any two vertices indicates that the corresponding operators do not commute. The figure visualizes the step-by-step coloring process used to partition this set based on commutation relations. In this simple illustrative example, both algorithms produce the same optimal coloring, yielding the color classes $\textbf{Orange} = \{XX, XI\}$, $\textbf{Green} = \{YY, YI\}$, and $\textbf{Blue} = \{ZZ, ZI\}$.

\subsubsection{Integer linear programming}
\label{sec:ILP}

To have a benchmark for smaller problem sizes, we use an exact algorithm for GC in addition to the heuristic ones. This algorithm uses \textit{integer linear programming} (ILP) to find a provably optimal solution to the GC problem by translating the coloring constraints and the objective into an integer linear program~\cite{anuj-INFORMS-96}. This method has exponential worst-case complexity. We note that ILP has been used in QST and other quantum-computing tasks before. For example, Ref.~\cite{li-pra-2017} used ILP to optimize local unitary operations to acquire tomographically complete data in an NMR-based quantum processor, but the reliance on local operations limits further efficiency gains. Similarly, in Ref.~\cite{verteletskyi-JCP2020}, clique-cover-based ILP was used to optimize measurements for the variational quantum eigensolver, with demonstrations for a set of molecular electronic Hamiltonians.

Here, we also frame the problem as an ILP-based minimum-clique-cover (MCC) task. The MCC task is to find the smallest number of cliques, i.e., fully connected subgraphs, needed to cover all the vertices of a given graph. This task is a natural complement to the GC problem. It can be represented by the complement graph $\tilde{\mathcal{G}}_{\mathcal{P}}$, where two vertices are connected if and only if they are not connected in the original graph $\mathcal{G}_{\mathcal{P}}$ as defined in \eqref{adj_mat}. Thus, in our case, the complement graph $\tilde{\mathcal{G}}_{\mathcal{P}} = (\mathcal{V}, \tilde{\mathcal{E}})$ simply represents the commutation graph: two vertices $v_i$, $v_j$ $\in \mathcal{V}$ are connected by an edge $e_{ij} \in \tilde{\mathcal{E}}$ if and only if the corresponding operators $P_i$ and $P_j$ commute, i.e., $[P_i, P_j] = 0$. Therefore, each clique corresponds to a color class that forms a subset of mutually commuting operators.

In the MCC-ILP approach, described with pseudocode in Algorithm~\ref{alg:ilp_mcc} in \appref{app:Pseudocode}, we start by constructing the commutation graph $\tilde{\mathcal{G}}_{\mathcal{P}}$. Next, we identify all maximal cliques in $\tilde{\mathcal{G}}_{\mathcal{P}}$ using the Bron--Kerbosch algorithm~\cite{bron-cacm-1973, hagberg-proc-2008, networkx_github}. After that, we build the ILP formulation using binary decision variables $x_i \in \{0,1\}$, where each variable corresponds to the selection of a clique. The objective is to minimize the total number of selected cliques, effectively minimizing the number of commuting-operator partitions. This minimization is subject to the constraint $\sum_{i : v \in C_i} x_i \geq 1 \:\: \forall v \in V$, which ensures that every vertex (i.e., operator) is included in at least one clique. The resulting ILP is solved using the \textit{coin-or branch and cut} (CBC) solver~\cite{cbc-solver1}, yielding an optimal collection of mutually commuting subsets of operators.

\begin{figure}
    \centering
    \includegraphics[width=\linewidth]{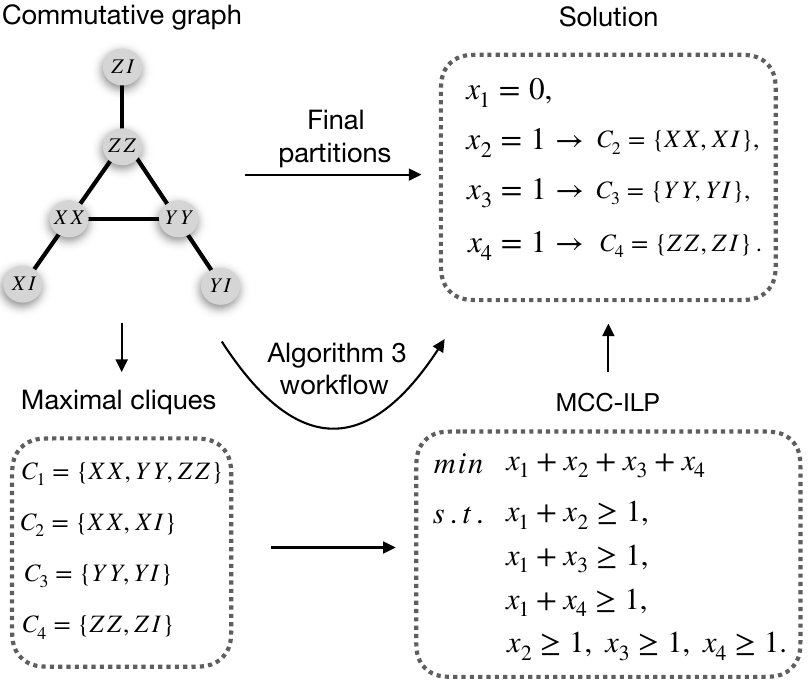}
    \caption{Illustration of the minimum-clique-cover (MCC) formulation of operator partitioning using an approach based on integer linear programming (ILP), demonstrated for the operator set $\mathcal{P} = \{ XI, YI, ZI, XX, YY, ZZ \}$ (the same as in \figref{fig:dsatur_rlf_coloring}). The procedure begins by constructing the corresponding commutation graph, followed by identification of maximal cliques. The MCC is then formulated as an ILP optimization task using Algorithm~\ref{alg:ilp_mcc} in \appref{app:Pseudocode}, and the final partitions are obtained by solving the resulting ILP.}
    \label{fig:mcc-ilp}
\end{figure}

To illustrate this method, we show in \figref{fig:mcc-ilp} how the MCC-ILP algorithm handles the same subset of two-qubit Pauli operators as in the example for DSATUR and RLF in \figref{fig:dsatur_rlf_coloring}:
$\mathcal{P} = \{ XI, YI, ZI, XX, YY, ZZ \}$.
We first construct the commutation graph $\tilde{\mathcal{G}}_{\mathcal{P}}$, which is the complement of the graph $\mathcal{G}_{\mathcal{P}}$ shown in Fig.~\ref{fig:dsatur_rlf_coloring}. Using the Bron--Kerbosch algorithm, we identify four maximal cliques from this commutation graph:
$C_1 = \{ XX, YY, ZZ \}$, 
$C_2 = \{ XX, XI \}$, 
$C_3 = \{ YY, YI \}$, and
$C_4 = \{ ZZ, ZI \}$.
Following Step 3 of Algorithm~\ref{alg:ilp_mcc} in \appref{app:Pseudocode}, we then formulate the ILP problem and solve it using the CBC solver. The optimal solution is the selection of the cliques $C_2$, $C_3$, and $C_4$, yielding the minimal (optimal) partition. 

Note that the number of variables in the objective function and the number of linear constraints in the ILP scales linearly as the number of maximal cliques and the number of vertices in the graph, respectively. To handle this growth better, the optimization process can be significantly accelerated by leveraging high-performance commercial solvers such as GUROBI or MOSEK~\cite{cvx-solvers}.


\subsubsection{Graph neural network}
\label{sec:GNN}

Another heuristic approach to solving the GC problem is using graph neural networks (GNNs). A GNN is a learning-based model designed to operate directly on graph-structured data. This design enables GNNs to tackle GC by learning to recognize structural patterns in graphs and producing color assignments with a single forward pass at inference time~\cite{velickovic-graphAttention-18, dai-combOpt-17}. Although GNNs have been used in quantum-information tasks such as quantum error correction~\cite{lange2025data, nink-ieee-2024, ryota-arxiv-2025}, their use for tomography-oriented experiment design is less explored.

Here, we apply a GNN to graph coloring (GC-GNN), in order to identify subsets of mutually commuting operators, by treating the coloring of $\mathcal{G_P}=(\mathcal{V},\mathcal{E})$ as a node-classification problem~\cite{hamilton-neurips2017-graphsage, zhou-aiopen-gnnreview}: each vertex $v\in\mathcal{V}$ is a training data, and its target label is the color (commuting group) assigned by an optimal or near-optimal method. The adjacency structure encodes non-commutation constraints, and through message passing, the network learns to extract both local and longer-range dependencies that are useful for predicting consistent color assignments~\cite{gilmer-NeuralMessagePassage-17}. Once trained, the model can generalize to unseen graphs of similar structure and rapidly output a complete partition of the operator set.

\begin{figure}
    \centering
    \includegraphics[width=\linewidth]{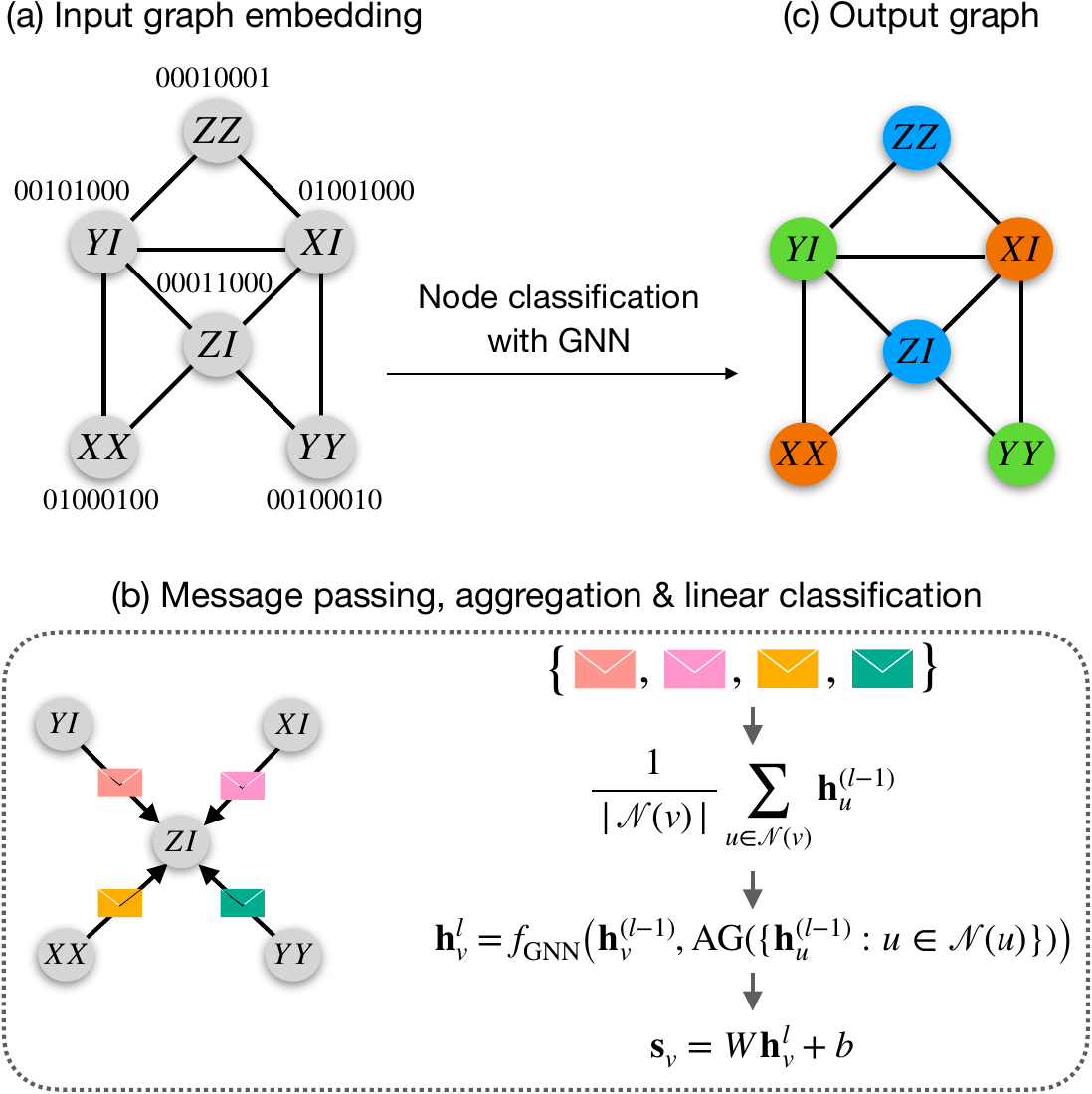}
    \caption{Graphical overview of the GNN-based GC framework. (a) Initial graph representation and node-feature embedding. (b) Internal GNN processing pipeline consisting of message passing, neighborhood aggregation, and linear classification. (c) Final GC solution, where nodes are assigned to color classes corresponding to mutually commuting operator partitions.}
    \label{fig:gnn}
\end{figure}

To illustrate the GC-GNN approach, which is detailed in Algorithm~\ref{alg:gnn_coloring} in \appref{app:Pseudocode} and further discussed in \appref{app:GNN}, we again consider the example of the two-qubit Pauli-operator subset $\mathcal{P}$ introduced in Fig.~\ref{fig:dsatur_rlf_coloring}. As shown in Fig.~\ref{fig:gnn}, each node $v$ is initialized with a simple feature vector (e.g., one-hot identity and/or basic structural descriptors, such as degree). The model then applies $l$ rounds of message passing.

At layer $\ell$, node $v$ aggregates information from its neighborhood $\mathcal{N}(v)$ using a permutation-invariant operator. Here, we use mean aggregation:
\begin{equation}
 \mathrm{AG}\mleft[\mleft\{\mathbf{h}_u^{(\ell-1)}\mright\}_{u\in\mathcal{N}(v)}\mright]
 = \frac{1}{|\mathcal{N}(v)|}\sum_{u\in\mathcal{N}(v)}\mathbf{h}_u^{(\ell-1)}.
\end{equation}
The node embedding is then updated as
\begin{equation}
 \mathbf{h}_v^{(\ell)}
 = f_{\rm GNN}\mleft(\mathbf{h}_v^{(\ell-1)},
 \mathrm{AG}\mleft[\mleft\{\mathbf{h}_u^{(\ell-1)}\mright\}_{u\in\mathcal{N}(v)}\mright]\mright),
\end{equation}
where $f_{\rm GNN}$ denotes a learnable update function (typically a linear transformation followed by a nonlinearity). After $l$ layers, $\mathbf{h}_v^{(l)}$ summarizes information from the $l$-hop neighborhood of $v$. A final linear classifier maps the embedding to scores over $k$ colors,
\begin{equation}
 \mathbf{s}_v = W \mathbf{h}_v^{(l)} + \mathbf{b}
 = \mleft[s_{v_1},\,s_{v_2},\,\dots,\,s_{v_k}\mright],
\end{equation}
and the predicted color is ${c}(v)=\arg\max_{k}s_{v_k}$. 

This procedure yields a full node-to-color assignment, and therefore a partition of the operator set into commuting measurement groups. The one-shot inference cost scales as $\mathcal{O}(l(|\mathcal{V}|+|\mathcal{E}|))$, making the approach efficient for rapid experiment scheduling once the model is trained~\cite{scarselli2009graph}; the main computational cost is the offline training stage.


\subsubsection{Spectral clustering}

The last algorithm we consider for GC is spectral clustering (SC), which leverages the eigenstructure of a graph's Laplacian matrix $L$ to partition nodes into clusters~\cite{ng-spectralClustering-02, strang2016ila, shi-IEEE-2000}. The eigenvectors of $L$ provide an effective way to assign real values to nodes such that connected nodes get similar values, and thus can be grouped by a clustering algorithm. 

In the QST context, we treat commutation as a similarity relation. Observables that commute form an affinity graph (the complement graph $\tilde{\mathcal{G}}_{\mathcal{P}}$ in \secref{sec:ILP}). Using SC, we cluster that graph to form joint measurement groups as described in Algorithm~\ref{alg:spectral_clustering} in \appref{app:Pseudocode}. In \appref{sec:SC}, we provide a complete description of how this SC algorithm works for the example of the two-qubit Pauli-operator subset $\mathcal{P}$ introduced in Fig.~\ref{fig:dsatur_rlf_coloring} and repeated for the other GC algorithms above. Below, we describe how the SC algorithm works in general.

In the SC algorithm, we begin by constructing an affinity matrix $S$ defined as
\begin{equation} \label{eq:sc}
    S_{ij} =
  \begin{cases}
    1, & \text{if}\;\;[P_i,P_j]=0 \;\; \forall i\neq j, \\
    0, & \text{if}\;\; i=j,  \\
    0, & \text{otherwise}.
  \end{cases} 
\end{equation}
Next, we form the degree matrix $D$---a diagonal matrix whose $i$th entry is the sum of row $i$ of $S$, representing how many commuting partners each observable has. Using these matrices, we compute the normalized Laplacian: $L \;=\; I \;-\; D^{-1/2}\,S\,D^{-1/2},$
where $I$ is the identity matrix. 

Small eigenvalues of $L$ correspond to node partitions that cross relatively few edges, implying that the graph can be split such that only a few commutation relations cross between clusters. We therefore perform an eigenvalue decomposition of $L$, extracting its $k$ smallest eigenvalues and corresponding eigenvectors. Stacking those $k$ eigenvectors columnwise into a matrix $\mathcal{M}$ produces an embedding, in which each row is a $k$-dimensional vector representing the node's projection. To ensure these embeddings are comparable, we normalize each row of $\mathcal{M}$ to unit length. 

Finally, we apply $k$-means clustering~\cite{MacQueen1967} to these normalized row vectors, such that nodes whose embeddings lie close together in this spectral space are grouped into the same cluster. Since the affinity matrix was built from commutation relations, each resulting cluster corresponds to a set of mutually commuting observables. 

The computational complexity of SC depends on the eigen-decomposition, which for dense graphs requires $O(n^3)$ time to compute all eigenvectors, where $n$ is the number of graph vertices (in our case, quantum operators)~\cite{Vargas-Lanczos-2020, vonluxburg-tutorialSC-07}.


\subsection{Designing and scheduling experiments for quantum state tomography}
\label{sec:scheduling}

The GC algorithms discussed in \secref{sec:gc-qst} provide heuristic (or, in the case of ILP, optimal) partitions of commuting operators. The heuristic partitioning is a fast way to find a small number of measurements to perform. However, once the partitioning is done, two more tasks remain before experiments are carried out and data can be collected: (i) constructing the measurement circuits and (ii) deciding on the order of the experiments. Here, we explain how we perform these two tasks.

First, to construct the measurement circuits, we note that once the operators are partitioned, the common eigenvectors of each commuting set can be computed. From these eigenvectors, we can construct change-of-basis unitary matrices, allowing measurements to be performed in the common eigenbasis and thereby enabling simultaneous measurement of all observables within a given set. In \appref{sec:qc}, we show one way to directly obtain these unitary matrices, using a greedy approach based on a binary symplectic representation of Pauli operators and symplectic Gaussian elimination. The resulting circuits only consist of elementary Clifford gates.

Second, to illustrate in which order to perform the experiments, consider a general system in a $d$-dimensional Hilbert space with a tomographically complete operator set $\mathcal{P} = \{ P_1, P_2,\cdots, P_{d^2} \}$. Suppose the given heuristic algorithm yields $m$ partitions $\{ \mathcal{C}_i \}_{i=1}^m$, each of cardinality $\mathcal{K}(\mathcal{C}_i)$,  satisfying $\sum_i \mathcal{K}(\mathcal{C}_i) = \mathcal{K}(\mathcal{P}) = d^2$. We prioritize QST experiments according to descending order of partition size, starting with the subset having the largest $\mathcal{K}(\mathcal{C}_i)$. This prioritization maximizes information gain and mitigates potential data redundancy, leading to more efficient state reconstruction than random sampling strategies. To demonstrate the benefits of this scheduling strategy, we compare its performance against conventional QST approaches that rely on random sampling from the set $\{ I, R_x, R_y \}^{\otimes N}$.


\section{Results and analysis}
\label{sec:results}

We now benchmark the GC algorithms described in \secref{methods} by applying them to several standard cases, including the partitioning of Pauli operators in multi-qubit systems (up to five qubits), Gell--Mann matrices in multi-qutrit systems (up to three qutrits), and hybrid systems with both qubits and qutrits. We focus on the solution optimality---measured by the number of commuting partitions---and the corresponding computational runtime. We further demonstrate the practical utility of these partitioning strategies for optimizing and scheduling QST experiments, and compare their performance against a conventional QST data-acquisition protocol. To be clear, we note again here that, in contrast to many approaches that primarily target improvements in classical post-processing, our focus is on optimizing the data-acquisition stage of QST by reducing and scheduling the required measurement settings.


\subsection{Partitioning}
\label{sec:partition-results}

\begin{figure*}
    \centering
     \includegraphics[width=\linewidth]{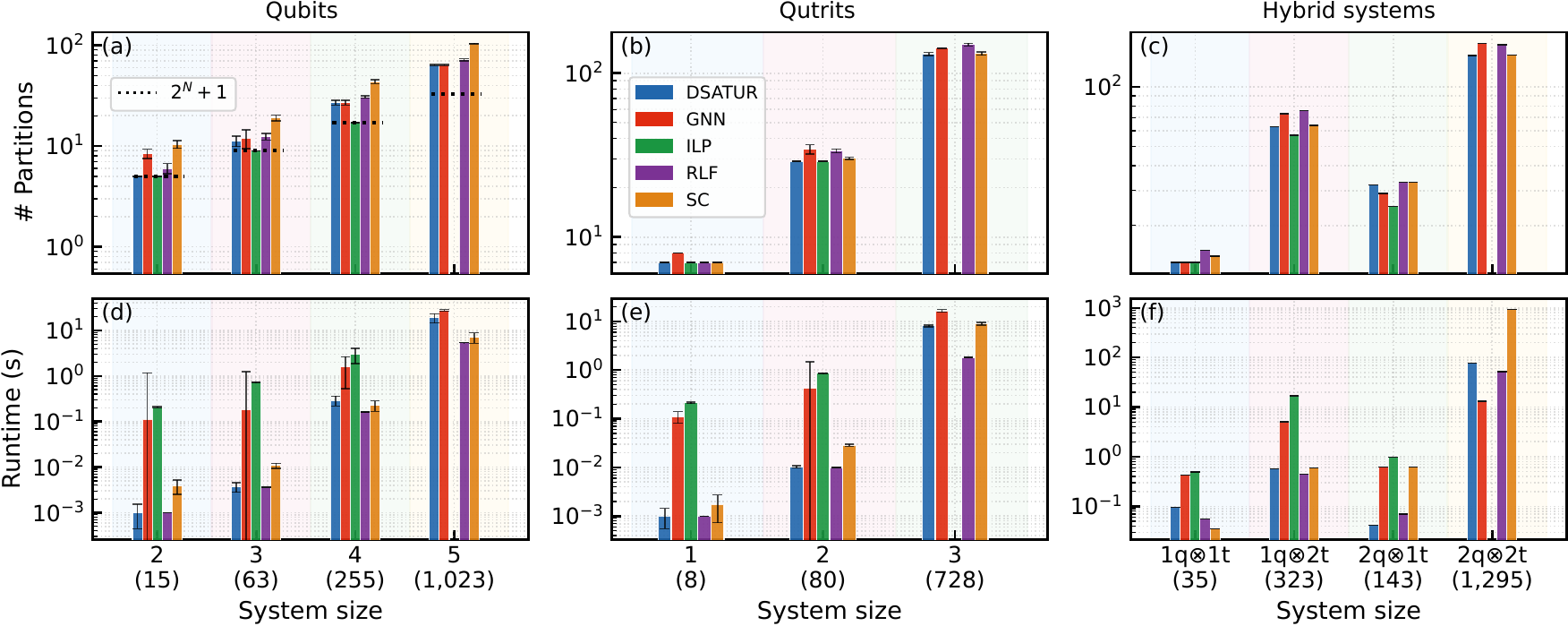}
    \caption{Benchmarking of graph-coloring (GC) algorithms for operator partitioning. 
    (a)--(c) The number of partitions obtained and
    (d)--(f) the corresponding computational runtime for five different GC algorithms (shown in different colors) across multi-qubit systems [first column, i.e., (a) and (d)], multi-qutrit systems [second column, i.e., (b) and (e)], and hybrid systems with both qubits and qutrits [third column, i.e., (c) and (f)]. In all panels, the x axis shows the system size (number of qubits and qutrits), with the total number of operators indicated in parentheses below each case. In panel (a), the dotted black line indicates the optimal reference value, $2^N + 1$ partitions, for the multi-qubit case.}
    \label{fig:Partition_count}
\end{figure*}

Figure~\ref{fig:Partition_count} summarizes the overall performance of the proposed GC methods for partitioning operator sets into mutually commuting subsets across different quantum systems. The first column [Figs.~\figpanelNoPrefix{fig:Partition_count}{a} and \figpanelNoPrefix{fig:Partition_count}{d}] presents results for multi-qubit systems, while the second [Figs.~\figpanelNoPrefix{fig:Partition_count}{b} and \figpanelNoPrefix{fig:Partition_count}{e}] and third columns [Figs.~\figpanelNoPrefix{fig:Partition_count}{c} and \figpanelNoPrefix{fig:Partition_count}{f}] show results for multi-qutrit and hybrid qubit-qutrit systems, respectively. In all panels of the first row, the y axis shows the number of obtained partitions, while the second row shows the corresponding computational runtime (in seconds) on a logarithmic scale. In \figpanel{fig:Partition_count}{a}, the black dashed line is the theoretical optimal partition count. Across all columns, the x axis indicates the respective system size, specified by the number of qubits (q) and qutrits (t), with the total number of operators given below each case (we have excluded the identity operator in all cases considered).

Across all panels in \figref{fig:Partition_count}, blue, red, purple, orange, and green bars represent the DSATUR, GNN, RLF, SC, and ILP algorithms, respectively. For the GNN, all reported statistics are from a model trained on the four-qubit graph; see \appref{app:GNN} for more details on the training and benchmarking of various GNN models.  

In all systems we tested, the ILP method found the optimal number of partitions. However, ILP quickly becomes intractable for larger problem instances, as indicated by the missing ILP results (due to too long runtime) for the five-qubit, three-qutrit, and $2\mathrm{q} \otimes 2\mathrm{t}$ cases in \figref{fig:Partition_count}. In contrast, all heuristic GC methods yielded solutions in significantly lower runtimes.

In the multi-qubit case, as shown in Figs.~\figpanelNoPrefix{fig:Partition_count}{a} and \figpanelNoPrefix{fig:Partition_count}{d}, the number of Pauli operators grows exponentially as $4^N - 1$ with the number of qubits $N$, increasing from $15$ for $N=2$ qubits to $1023$ for $N=5$ qubits. For a complete $N$-qubit Pauli operator set, the theoretical optimum number of mutually commuting partitions is $2^N+1$~\cite{zeilinger-pra-2002, reggio-pra-2024}, corresponding to $5$, $9$, $17$, and $33$ measurement settings for $N=2,3,4$, and $5$, respectively. 

While the ILP approach attained this optimum up to $N=4$, all heuristic methods produced strongly compressed commuting partitions across all system sizes. For instance, the DSATUR and GNN approaches each produced solutions with $25$ partitions for $N=4$ in a matter of seconds. In the five-qubit case, where ILP took too long to run, all heuristic approaches continued to generate valid commuting partitions for the complete $1023$-operator set, with DSATUR and GNN producing on average $63$ partitions in tens of seconds. Among the heuristic approaches, DSATUR consistently performs best or equal to the GNN in terms of the number of partitions, and it does so in a shorter runtime than the GNN. Note that the GNN runtime here is the inference time; the training time is about one minute. See \appref{app:GNN} for further details on the GNN training. The SC algorithm consistently yields larger partition counts, since it groups vertices according to spectral similarity of the graph Laplacian rather than explicitly optimizing the chromatic number.

Similarly, for the multi-qutrit systems shown in Figs.~\figpanelNoPrefix{fig:Partition_count}{b} and \figpanelNoPrefix{fig:Partition_count}{e}, we consider the complete set of generalized Gell--Mann operators. The size of this $N$-qutrit operator set thus grows as $9^N-1$. In the general qutrit case, the optimal number of partitions is not known, so we treat the ILP results as reference optimal solutions whenever they can be obtained.

For the simple single-qutrit case, all methods produce nearly identical partition counts. For the two-qutrit case, consisting of $80$ operators of dimension $9 \times 9$, the ILP method produces a solution with $29$ partitions in one second. The heuristic methods yield closely comparable results in about two orders of magnitude faster runtime than ILP, with DSATUR producing 29 partitions, GNN 32, RLF 33, and SC producing 30. 

In the three-qutrit case, the operator set increases to $728$ observables, each of dimension $27 \times 27$. Here, the ILP method does not converge within a practical runtime in our implementation, while all heuristic approaches continue to generate valid commuting partitions in about ten seconds or less. These qutrit results demonstrate that our GC-based partitioning framework extends naturally beyond multi-qubit Pauli systems and remains effective for higher-dimensional Hilbert spaces.

For the hybrid systems with both qubits and qutrits shown in Figs.~\figpanelNoPrefix{fig:Partition_count}{c} and \figpanelNoPrefix{fig:Partition_count}{f}, the operator sets contain 35, 323, 143, and 1295 observables for the $1\mathrm{q}\otimes1\mathrm{t}$, $1\mathrm{q}\otimes2\mathrm{t}$, $2\mathrm{q}\otimes1\mathrm{t}$, and $2\mathrm{q}\otimes2\mathrm{t}$ configurations, respectively. These systems have Hilbert-space dimensions of 6, 18, 12, and 36.

For the smallest hybrid system, $1\mathrm{q}\otimes1\mathrm{t}$, all methods yield comparable partition counts in under one second.
For the $1\mathrm{q}\otimes2\mathrm{t}$ system, the ILP method produces a solution with $57$ partitions, while DSATUR, GNN, RLF, and SC produce $63$, $73$, $76$, and $64$ partitions, respectively. The largest hybrid system considered in this work is the $2\mathrm{q}\otimes2\mathrm{t}$ case. For this system, the ILP solver does not converge within a practical runtime in our implementation, while all heuristic approaches remain computationally tractable and continue to generate valid commuting partitions. In particular, DSATUR and SC do best by producing 144 and 145 partitions, respectively, which took less than \SI{100}{\second} for DSATUR, but around \SI{1000}{\second} for SC. The GNN and RLF methods were even faster than DSATUR, but produced solutions of lower quality.

Taking stock of all the results in \figref{fig:Partition_count}, we see that our GC-based partitioning framework naturally extends from qubits to both qutrits and composite quantum systems with mixed local dimensions. It thus remains effective beyond the algebraic structure of standard multi-qubit Pauli operator sets. We also note that, among the heuristic algorithms, DSATUR seems to provide the best trade-off between finding few partitions and doing so in a short time. However, the GNN approach also performed well, and may be preferable in some cases where inference time matters more than training time and where training can be geared towards a particular system.


\subsection{Quantum state tomography}
\label{sec:qst-results}

The operator partitions obtained from the various GC methods can be further leveraged to design efficient QST measurement schedules. To maximize the information extracted at each experimental step, the mutually commuting subsets are ordered according to their cardinality (see \secref{sec:scheduling}), and measurements are performed from the largest to the smallest subset. We now test the performance of this method for the case of qubits.

\begin{figure*}
    \centering
    \includegraphics[width=1\linewidth]{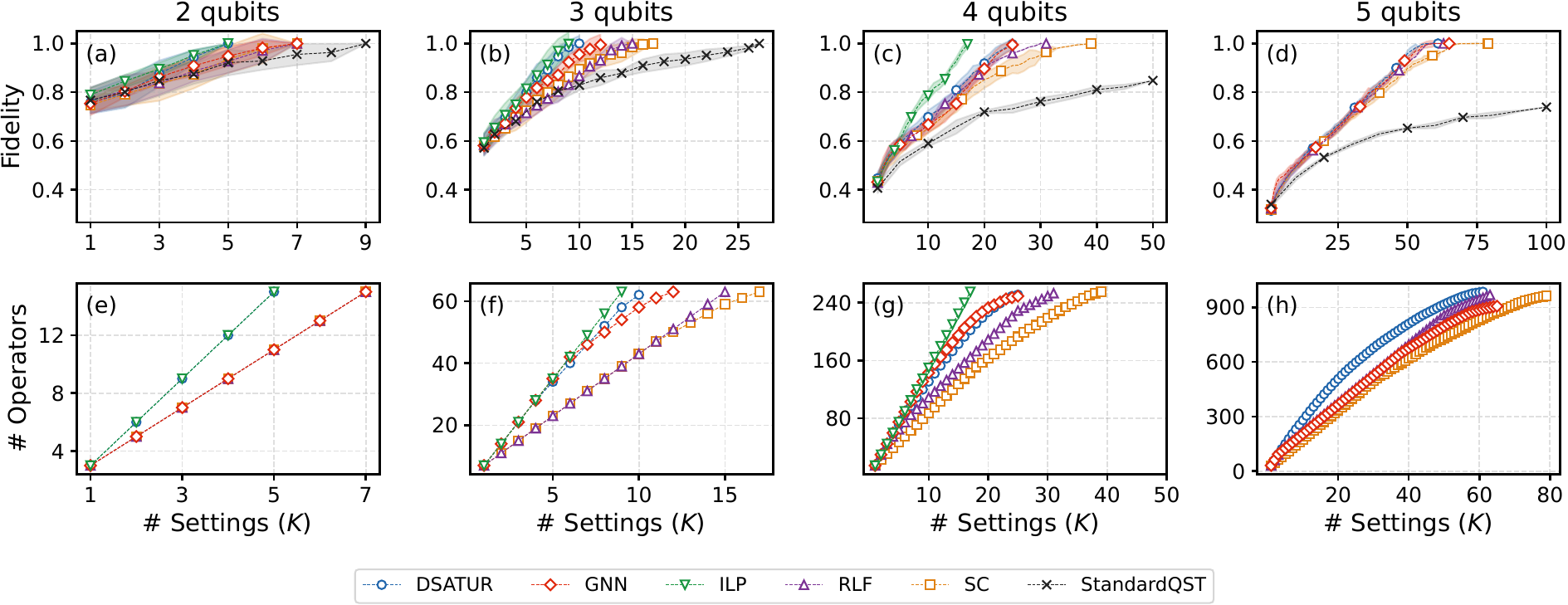}
    \caption{Benchmarking quantum state tomography with measurement schedules based on graph-coloring (GC) algorithms.
    (a)--(d) Reconstruction fidelity $F$ and
    (e)--(h) total number of Pauli operators measured, as a function of the number of GC-based scheduled measurement settings for two [(a), (e)], three [(b), (f)], four [(c), (g)], and five [(d), (h)] qubits. In the first row, the GC-based scheduled measurement settings (DSATUR: blue circles; GNN: red diamonds; ILP: green triangles; RLF: purple triangles; SC: orange squares) are compared with standard QST experiments (black crosses) sampled from $\{ I, R_x, R_y \}^{\otimes N}$. } 
    \label{scheduling}
\end{figure*}

In \figref{scheduling}, we show the results of these tests for full QST using GC-derived measurement schedules for multi-qubit states of up to five qubits. In all panels of the figure, the x axis shows the total number of measurement settings used for density-matrix reconstruction. These settings are progressively incorporated according to the descending cardinality of the commuting subsets generated by each GC method. The corresponding color coding is the same as in \figref{fig:Partition_count}: DSATUR (blue), SC (orange), RLF (purple), GNN (red), and ILP (green). 

The first row of Fig.~\ref{scheduling} shows the reconstruction performance achieved by the different measurement schedules for $N=2, 3, 4$, and 5 qubits. Here, the y axis is the Uhlmann--Jozsa state fidelity $F$~\cite{jozsa-jmp-1994} between the reconstructed and target density matrices. The average fidelity values for the different methods are shown as solid colored curves together with the corresponding standard deviation (shaded regions). These numbers are calculated from reconstructions of 15 randomly generated mixed $N$-qubit target density matrices, generated in QuTiP~\cite{Johansson2012, Johansson2013QuTiP2, Lambert2026} using \texttt{rho\_true = rand\_dm(2**N, density=1.0, dims=[[2]*N, [2]*N])}. To benchmark the GC-based measurement schedules, we also include results obtained using a conventional QST protocol (black curve), in which measurement settings are randomly sampled from $\{I, R_x, R_y\}^{\otimes N}$.

The second row of Fig.~\ref{scheduling} provides a complementary view of the measurement efficiency. In these panels, we show the cumulative number of observables covered as a function of the total number of measurement settings for the GC methods. The measured observables are subsequently used for the state reconstructions reported in the corresponding panels of the first row.
After each additional measurement setting is incorporated, the quantum state is reconstructed using a least-squares estimator (convex optimization using CVX~\cite{cvx-solvers}). This procedure enables a direct comparison of the rate at which information is accumulated by the different scheduling strategies and provides insight into how rapidly high-fidelity state reconstruction can be achieved using various methods.

Although standard QST data-acquisition methods are experimentally straightforward to implement, they generally require a substantially larger number of measurement settings to collect sufficiently informative and non-redundant data for accurate state reconstruction. In contrast, the proposed GC-based measurement schedules prioritize settings according to the size of their corresponding commuting partitions, thereby maximizing the amount of information extracted from the earliest measurements. This advantage can be seen in all the panels in the first row of \figref{scheduling}, where high-fidelity reconstruction is achieved much faster using the GC-derived measurement settings than with standard QST. For instance, for the four-qubit system shown in \figpanel{scheduling}{c}, the GC-heuristic-based schedulers GNN and DSATUR achieve a reconstruction fidelity close to 1 in only 25 measurement settings, whereas the standard QST protocol attains a fidelity below 0.75 using the same number of settings; the RLF and SC methods also clearly outperform standard QST here.

Among the GC methods in the four-qubit case, the ILP-based measurements achieve nearly perfect state reconstruction using only 17 measurement settings, which coincides with the theoretical optimum of $2^4+1=17$ settings to cover the complete set of four-qubit Pauli operators, as discussed in \secref{sec:partition-results}. The origin of this improvement can also be observed in the corresponding complementary plot in \figpanel{scheduling}{g} for the cumulative number of measured observables. While the heuristic GC methods require approximately 30 partitions to achieve complete operator coverage, the conventional QST scheme requires $3^4 = 81$ measurement settings to do the same. Consequently, among all methods, the ILP-based scheduling attains full informational completeness significantly earlier, enabling high-fidelity state reconstruction with considerably fewer measurements. However, as we have seen in \secref{sec:partition-results}, ILP is not feasible to use for larger system sizes due to its long runtime.

A common feature for the GC methods across first row of \figref{scheduling} is the rapid initial increase in reconstruction fidelity. Since the measurement settings obtained from these methods are ordered according to the cardinality of their commuting partitions, the first few experiments correspond to the most information-rich subsets of observables. Consequently, a significant fraction of the state information is recovered early in the measurement process. For example, in the five-qubit case, all the GC-based scheduled measurement settings achieve an average reconstruction fidelity exceeding 0.9 after the first 50 settings, despite requiring more than 60 partitions for complete operator coverage. At the same time, the standard QST only achieves 0.65 state fidelity with same number of measurement settings. 

From a practical perspective, this priority-based measurement schedule enables experimentalists to maximize information gain at each stage of the tomography procedure and to make informed decisions during data acquisition. Once a target reconstruction fidelity has been reached, the remaining measurements may be omitted, thereby reducing experimental overhead and measurement time. When complete state characterization is required, the proposed framework still provides a full, but relatively small, set of measurement settings that guarantees informational completeness. Indeed, in our simulations with ideal, noise-free measurement data, performing measurements over all scheduled partitions yields essentially exact state reconstruction, achieving infidelities $1-F < 10^{-7}$.


\section{Conclusion and outlook}
\label{sec:conclusion}

We developed a general graph-coloring (GC)-based framework for optimization and scheduling of quantum state tomography (QST) experiments through the partitioning of arbitrary operator sets into mutually commuting subsets. These subsets correspond to groups of observables that can be measured simultaneously, thereby reducing experimental complexity while preserving informational completeness for the state reconstruction.

Importantly, our framework is not restricted to multi-qubit Pauli operators, but extends naturally to arbitrary operator sets and arbitrary-dimensional Hilbert spaces, including subsets of the Pauli operators, multi-qutrit systems, as well as hybrid architectures with both qubits, qutrits, and other qudits. This versatility positions our framework as a unified methodology for tomographic optimization across a broad range of quantum platforms. In this general setting, it has thus far been hard to do partitioning of operator sets well, since the ways to partition them grows combinatorially, rendering brute-force and other exact methods infeasible already for systems with just a few qubits or qudits.

By casting the measurement-setting problem as a GC task, we were able to access efficient heuristic algorithms developed for that task and apply them to operator partitioning instead. In this work, we tested the heuristic algorithms degree of saturation (DSATUR), recursive largest first (RLF), spectral clustering (SC), and graph neural networks (GNNs). We benchmarked these algorithms against the exact method of integer linear programming (ILP) for small system sizes and against standard QST measurement settings.

In numerical studies on systems containing up to five qubits, three qutrits, or two qubits and two qutrits, we found that the heuristic GC algorithms substantially reduced the number of distinct measurement settings (compared to standard QST) required for complete state reconstruction. The resulting partitions were consistently close to the optimal solutions whenever exact benchmarks were available. Furthermore, the heuristic algorithms only needed seconds to find good partitions even when the system size was such that the exact method of ILP took too long to run. Weighing both the performance in terms of measurement settings and runtime, our studies suggested that DSATUR was the heuristic algorithm that performed best overall, with GNN as a promising runner-up.

Based on the operator partitioning, we also introduced a scheduling strategy for selecting which experiments to run: we order commuting subsets according to their cardinality, ensuring that the most informative measurements are performed first. We found that this information-driven prioritization accelerates the accumulation of tomographic information compared to standard QST and leads to a rapid increase in reconstruction fidelity during the early stages of data acquisition.

Our proposed framework thus enables high-accuracy QST with substantially fewer measurement settings than standard approaches, while retaining the ability to perform complete state reconstruction when required. Beyond reducing experimental overhead, the scheduling paradigm naturally supports adaptive stopping criteria, allowing measurements to be terminated once a desired fidelity threshold has been reached. These features highlight the practical advantages of GC-based experimental design and establish it as a promising tool for scalable and resource-efficient QST.

As quantum processors (and other quantum technologies, e.g., quantum sensors and quantum communication networks) continue to increase in size and complexity, this heuristic approach therefore provide a practical pathway toward (more) scalable characterization. This approach to operator partitioning can also have other important applications, including in quantum simulation and certain quantum algorithms, where efficient measurements of commuting observables play a central role.  

Despite the advantages offered by GC-based measurement scheduling, its practical implementation on quantum hardware still has challenges. In particular, there is currently no general and efficient procedure for taking an arbitrary set of mutually commuting observables, produced by the operator partitioning, and systematically construct and implement a change-of-basis transformation enabling the corresponding measurement using a universal gate set. Although we in this article provided a procedure for synthesizing such measurement circuits from only elementary Clifford gates, we believe there is still room for improvement here. Even for multi-qubit systems, implementing measurements in the common eigenbasis of an arbitrary set of mutually commuting Pauli operators remains a nontrivial task and represents an important direction for future research. Here, it is likely important to optimize both the circuit depth and the amount of required entangling operations. Addressing this challenge would help extend the practical utility of GC-based measurement strategies beyond QST, benefiting both quantum simulation and some quantum algorithms.

Looking ahead, one direction for future research would be to test the performance of the heuristic GC algorithms for larger systems, to get a better understanding of which algorithm scales best. Here, we believe it is particularly interesting to investigate GNNs more deeply. Once trained on representative operator-commutation graphs, the GNN learns the structural characteristics of the partitioning problem and can subsequently predict efficient measurement groupings for previously unseen quantum systems in a short inference time. Indeed, the GNN seems to generalize surprisingly well to larger systems from being trained on just a small example. Across the systems we tested, the GNN consistently produced partition counts that were close to those obtained by exact or heuristic GC methods, while maintaining low computational overhead (milliseconds to seconds; for the largest system tested, the GNN runtime was the lowest of all methods). By combining data-driven generalization through machine learning with efficient operator partitioning, the GNN-based framework potentially provides a practical and scalable solution for measurement scheduling in quantum systems whose size and complexity would otherwise render exhaustive or exact approaches impractical.

Finally, another natural extension of the GC framework is its application to quantum process tomography (QPT). Unlike QST, where the optimization focuses on measurement settings alone, QPT requires a joint optimization of both probe input states and measurement configurations, especially in the case of ancilla-assisted QPT~\cite{marco-arxiv-2018, xue-prl-2022, lu-anderphys-2022}. Extending graph-theoretic partitioning and scheduling strategies to this more general setting could lead to significant reductions in the experimental complexity for quantum process characterization.


\begin{acknowledgments} 

We acknowledge support from the Knut and Alice Wallenberg Foundation through the Wallenberg Centre for Quantum Technology (WACQT) and the Horizon Europe programme HORIZON-CL4-2022-QUANTUM-01-SGA via the project 101113946 OpenSuperQPlus100. AFK is also supported by the Swedish Foundation for Strategic Research (grant numbers FFL21-0279 and FUS21-0063). GitHub Copilot student (Claude Haiku) and GPT 5.5 were used to assist with code development and refinement, generation of plotting code, and linguistic refinement of the manuscript. All LLM-assisted content was reviewed and verified by the authors, who take full responsibility for the final work.  

\end{acknowledgments}


\appendix


\section{Pseudocode for graph-coloring algorithms}
\label{app:Pseudocode}

Here, we present pseudocode for the five GC algorithms we employ in this article: Algorithm~\ref{dsatur} is DSATUR, Algorithm~\ref{rlf} is RLF, Algorithm~\ref{alg:ilp_mcc} is ILP for MCC, Algorithm~\ref{alg:gnn_coloring} is the GNN, and Algorithm~\ref{alg:spectral_clustering} is SC.

\begin{algorithm}[H]
\caption{Degree of saturation}
\label{dsatur}
\begin{algorithmic}[1]
\Require Graph $\mathcal{G_P} = (\mathcal{V}, \mathcal{E})$
\State Initialize all vertices as uncolored
\While{there are uncolored vertices}
    \State Select the uncolored vertex with the highest saturation degree
    \If{multiple vertices have the same saturation degree}
        \State Select the one with the highest degree (most neighbors)
    \EndIf
    \State Assign the smallest available color not used by adjacent colored vertices
\EndWhile
\end{algorithmic}
\end{algorithm}

\begin{algorithm}[H] 
\caption{Recursive largest first}
\label{rlf}
\begin{algorithmic}[1]
\Require Graph $\mathcal{G_P} = (\mathcal{V}, \mathcal{E})$
\State Initialize all vertices as uncolored
\While{there are uncolored vertices}
    \State Select the uncolored vertex with the highest degree
    \State Initialize a new color class with this vertex
    \ForAll{uncolored vertices not adjacent to any vertex in the current color class}
        \State Add them to the current color class
    \EndFor
    \State Assign a new color to all vertices in the current color class
    \State Remove vertices in current color class from the graph
\EndWhile
\end{algorithmic}
\end{algorithm}

\begin{algorithm}[H]
  \caption{Integer linear programming for minimum clique cover}
  \label{alg:ilp_mcc}
  \begin{algorithmic}[1]
    \Require Commutativity graph $\tilde{\mathcal{G}}_\mathcal{P} = (\mathcal{V}, \tilde{\mathcal{E}})$
    \State Find all maximal cliques \( \mathcal{C} = \{C_1, C_2, \dots, C_k\} \)
    \State Define binary variables \( x_i \in \{0,1\} \) for each \( C_i \in \mathcal{C} \)
    \State Define ILP:
    \[
        \min \sum_{i=1}^{k} x_i
    \]
    \[
        \text{s.t. } \sum_{i : v \in C_i} x_i \geq 1 \quad \forall v \in V
    \]
    \State Solve the ILP
    \State \textbf{Return} Selected cliques \( \{ C_i \mid x_i = 1 \} \)
    \State Assign colors to selected cliques
  \end{algorithmic}
\end{algorithm}

\begin{algorithm}[H]
\caption{Coloring with a graph neural network}
\label{alg:gnn_coloring}
\begin{algorithmic}[1]
\State \textbf{Input:}  $\mathcal{G_P}$, pretrained GNN with $l$ layers
\State \textbf{Output:} A color label for each node
\medskip
\State \textbf{Initialize node features}
\State \quad Assign each node an initial embedding (e.g., one-hot identity, degree, or other simple structural features)
\medskip
\State \textbf{Message passing (repeat for each of the $l$ layers)}
\State \quad For each node:
\State \quad\quad Aggregate embeddings from neighboring nodes
\State \quad\quad Update the node embedding using the GNN layer (linear map + nonlinearity)
\medskip
\State \textbf{Color prediction}
\State \quad For each node:
\State \quad\quad Apply the GNN's final output layer to its feature to produce a score for each possible color
\State \quad\quad Choose the color with the highest score and assign it to the node
\medskip
\State \textbf{Return} the set of node to color assignments
\medskip
\State \textbf{Notes:}
\State \quad Each GNN layer is typically a small neural network block (linear layer + activation).
\State \quad The final output layer maps node features to a score for each color.
\end{algorithmic}
\end{algorithm}

\begin{algorithm}[H]
\caption{Spectral clustering via affinity graph}
\label{alg:spectral_clustering}
\begin{algorithmic}[1]
\Require Graph $\mathcal{G_P}$ where edges link non-commuting observables, desired cluster count $k$
\Ensure A cluster label for each node (defines commuting measurement groups)
\medskip
\State \textbf{Build the affinity (commutation) matrix}
\State \quad For each unordered pair of distinct vertices $v_i,v_j \in \mathcal{V}$:
\State \quad Let $e_{ij}=\{v_i,v_j\}$. If $e_{ij}\notin\mathcal{E}$, equivalently $[P_i,P_j]=0$. 
\State \quad Set $S_{ij}=S_{ji}=1$ (they commute)
\State \quad Set diagonal entries $S_{ii}=0$
\medskip
\State \textbf{Compute degrees}
\State \quad For each node $i$, let $\mathrm{deg}_i = \sum_j S_{ij}$
\medskip
\State \textbf{Form normalized Laplacian}
\State \quad Construct $D$ as the diagonal matrix of degrees
\State \quad Compute $L = I - D^{-1/2}\,S\,D^{-1/2}$
\medskip
\State \textbf{Spectral embedding}
\State \quad Compute the $k$ eigenvectors of $L$ corresponding to its $k$ smallest eigenvalues
\State \quad Stack these eigenvectors column-wise into an $\mathcal{|V|}\times k$ matrix $\mathcal{M}$
\medskip
\State \textbf{Row normalization}
\State \quad For each row $i$ of $\mathcal{M}$, perform normalization so it has length 1
\medskip
\State \textbf{Cluster assignment}
\State \quad Run $k$-means clustering on the rows of $\mathcal{M}$ to assign each node to one of $k$ clusters
\medskip
\State \Return The cluster labels for all nodes
\end{algorithmic}
\end{algorithm}

\section{Training and benchmarking of graph neural networks}
\label{app:GNN}

All GNN models used in this study are supervised message-passing networks (see \secref{sec:GNN} and Algorithm~\ref{alg:gnn_coloring} in \appref{app:Pseudocode}) trained on graphs generated using DSATUR-based colorings.
All GNN results reported in Figs.~\ref{fig:Partition_count} and \ref{scheduling} were obtained using a model trained on the four-qubit commutation graph. 

The inference time of the GNN is not only lower than that of ILP, but it is relatively better than other heuristic methods within the tested regime. Here, one also needs to remember the cost associated with training and model initialization, which other methods do not have. This distinction between offline training and online inference is essential: the computationally intensive training phase is performed once, after which the trained model can be reused to generate operator partitions for unseen graph instances with minimal overhead.

\begin{figure*}
    \centering
    \includegraphics[width=\linewidth]{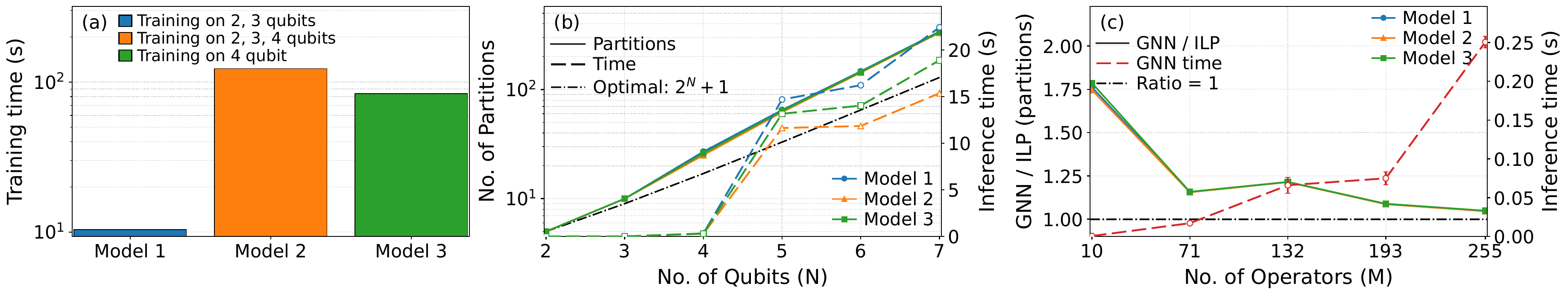}
    \caption{Learning and inference behavior of GNN-based operator partitioning.
  (a) Training clock time per model. The bars correspond to the trained models and are colored by the training scenario (number of qubits used).
  (b) Generalization across target system size $N$. The left y axis shows predicted partition count. The right y axis shows inference time in seconds. 
  Error bars denote standard deviation across multiple runs, but are too small to be visible in the figure.
  The black dotted line shows the optimal reference $2^{N}+1$. (c) Comparison of GNN and ILP performance (left y axis) for four-qubit operator subsets with $M$ operators in each. Operators are sampled over 5 random seeds. The right y axis reports GNN inference time only (red dashed line; data markers show the mean~$\pm$~standard deviation across runs).}
    \label{fig:GNN_Output}
\end{figure*}

To better assess the scalability and generalization capability for the GNN approach, we here further analyze its training and inference behavior. In \figpanel{fig:GNN_Output}{a}, we show the training times for three GNN models. Model 1 is trained on graphs corresponding to two- and three-qubit systems, Model 2 is trained on two-, three-, and four-qubit graphs, while Model 3 is trained exclusively on four-qubit data. As expected, training time increases with graph size and density, reflecting the higher computational cost of learning from larger commutation structures. Once trained, all models are applied to previously unseen commutation graphs via a single forward inference pass.

In \figpanel{fig:GNN_Output}{b}, we demonstrate that the predicted number of partitions remains relatively close to the reference optimum $2^N+1$ across the tested numbers of qubits in the main text (up to five), and beyond that also up to seven qubits. For instance, using Model 2 for the five-qubit case yields 62 predicted partitions with an inference time of approximately $1.5$s. Importantly, the trained models can also be applied to arbitrary operator subsets beyond informationally complete sets. 

Model 2, having been exposed to larger training instances, exhibits slightly improved performance for higher qubit numbers compared to Model 1, but both models remain closely aligned on the tested benchmarks. Indeed, it is surprising how well Model 1 performs given its limited training data. This performance suggests that the GNN is good at generalizing to larger problem sizes for GC, which is promising for applying this method to even larger systems.

In \figpanel{fig:GNN_Output}{c}, we further compare GNN predictions with ILP solutions for randomly sampled four-qubit operator subsets of varying size $M$. The ratio of GNN to ILP partition counts is close to two for smaller subsets, but reduces to close to unity for larger subsets. Further investigation is warranted to understand why the GNN performs worse for smaller subsets; it may be due to the training data. In all cases, the GNN inference time is negligible relative to the time taken for exact optimization. 

Overall, the GNN-based approach combines strong partitioning performance with promising scalability, making it a practical tool for decomposing arbitrary operator sets into commuting subsets. Unlike classical heuristics, the model learns structural features of the underlying graph directly from data, including subtle commutation patterns that enable generalization beyond the training distribution. This provides an example of how machine learning can be effectively integrated into quantum information processing workflows.


\section{Example demonstration for spectral clustering}
\label{sec:SC}

Here, we illustrate how the spectral-clustering method works by applying it to the same example we used for DSATUR and RLF in \figref{fig:dsatur_rlf_coloring}, for ILP in \figref{fig:mcc-ilp}, and for GNN in \figref{fig:gnn}. We consider the following subset of two-qubit Pauli operators:
\begin{equation}\label{eq:B2}
    \mathcal{P}=\{XX,YY,ZZ,XI,YI,ZI\}.
\end{equation}

Following Algorithm~\ref{alg:spectral_clustering} in \appref{app:Pseudocode}, we first construct the commutation-affinity matrix $S$ using \eqref{eq:sc}:
\begingroup
\small
\begin{equation}
S=
\mleft[
\begin{array}{c|cccccc}
 & XX & YY & ZZ & XI & YI & ZI \\
\hline
XX & 0 & 1 & 1 & 1 & 0 & 0 \\
YY & 1 & 0 & 1 & 0 & 1 & 0 \\
ZZ & 1 & 1 & 0 & 0 & 0 & 1 \\
XI & 1 & 0 & 0 & 0 & 0 & 0 \\
YI & 0 & 1 & 0 & 0 & 0 & 0 \\
ZI & 0 & 0 & 1 & 0 & 0 & 0
\end{array}
\mright].
\end{equation}
\endgroup
It follows that the degree matrix $D$ is
\begin{equation}
    D=\operatorname{diag}(3,3,3,1,1,1).
\end{equation}
Using $S$ and $D$, we then compute the normalized graph Laplacian $L$ according to step 10 in Algorithm~\ref{alg:spectral_clustering}:
\begingroup
\small
\begin{equation}
L=
\left[
\begin{array}{cccccc}
1 & -\frac{1}{3} & -\frac{1}{3} & -\frac{1}{\sqrt{3}} & 0 & 0\\
-\frac{1}{3} & 1 & -\frac{1}{3} & 0 & -\frac{1}{\sqrt{3}} & 0\\
-\frac{1}{3} & -\frac{1}{3} & 1 & 0 & 0 & -\frac{1}{\sqrt{3}}\\
-\frac{1}{\sqrt{3}} & 0 & 0 & 1 & 0 & 0\\
0 & -\frac{1}{\sqrt{3}} & 0 & 0 & 1 & 0\\
0 & 0 & -\frac{1}{\sqrt{3}} & 0 & 0 & 1
\end{array}
\right].
\end{equation}
\endgroup
The eigenvalues of $L$ are
\begin{equation}
\begin{aligned}
\lambda(L)=
\Bigg\{&
0,\,
\frac{7-\sqrt{13}}{6},\,
\frac{7-\sqrt{13}}{6},\,
\frac{4}{3},\\
&
\frac{7+\sqrt{13}}{6},\,
\frac{7+\sqrt{13}}{6}
\Bigg\}.
\end{aligned}
\end{equation}

For a desired cluster count of $k=3$, spectral clustering uses the eigenvectors corresponding to the three smallest eigenvalues. The eigenvector for the zero eigenvalue is
\begin{equation}
    \boldsymbol{\phi}_{1}
    =
    \frac{1}{2\sqrt{3}}
    \begin{pmatrix}
    \sqrt{3}\\
    \sqrt{3}\\
    \sqrt{3}\\
    1\\
    1\\
    1
    \end{pmatrix}.
\end{equation}
The next two eigenvectors span the degenerate eigenspace corresponding to
\begin{equation}
    \lambda_2=\lambda_3=\frac{7-\sqrt{13}}{6}, 
\end{equation}
which has the orthonormal basis vectors
\begin{equation}
\begin{aligned}
\boldsymbol{\phi}_{2}
&=
\frac{1}{n_2}
\begin{pmatrix}
1 & -1 & 0 & \beta & -\beta & 0
\end{pmatrix}^{T},\\
\boldsymbol{\phi}_{3}
&=
\frac{1}{n_3}
\begin{pmatrix}
1 & 1 & -2 & \beta & \beta & -2\beta
\end{pmatrix}^{T} ,
\end{aligned}
\end{equation}
where 
\begin{equation}
\begin{gathered}
    \beta=\frac{\sqrt{39}+\sqrt{3}}{6},\\[1mm]
    n_2=\sqrt{2(1+\beta^2)},\\[1mm]
    n_3=\sqrt{6(1+\beta^2)}.
\end{gathered}
\end{equation}

The spectral embedding matrix is then constructed by stacking these eigenvectors column-wise:
\begin{equation}
    \mathcal{M}
    =
    \begin{pmatrix}
    | & | & |\\
    \boldsymbol{\phi}_{1} & \boldsymbol{\phi}_{2} & \boldsymbol{\phi}_{3}\\
    | & | & |
    \end{pmatrix}.
\end{equation}
Explicitly, the rows of $\mathcal{M}$ are
\begingroup
\small
\begin{equation}
\begin{array}{c|ccc}
 & \phi_1 & \phi_2 & \phi_3\\
\hline
XX & \frac{1}{2} & \frac{1}{n_2} & \frac{1}{n_3}\\
YY & \frac{1}{2} & -\frac{1}{n_2} & \frac{1}{n_3}\\
ZZ & \frac{1}{2} & 0 & -\frac{2}{n_3}\\
XI & \frac{1}{2\sqrt{3}} & \frac{\beta}{n_2} & \frac{\beta}{n_3}\\
YI & \frac{1}{2\sqrt{3}} & -\frac{\beta}{n_2} & \frac{\beta}{n_3}\\
ZI & \frac{1}{2\sqrt{3}} & 0 & -\frac{2\beta}{n_3}
\end{array}.
\end{equation}
\endgroup

The $i$th row of $\mathcal{M}$, denoted by $\mathbf{m}_i$, gives the spectral-coordinate vector of vertex $v_i$. Before applying $k$-means clustering, each row is normalized as
\begin{equation}
    \widehat{\mathbf{m}}_i
    =
    \frac{\mathbf{m}_i}{\|\mathbf{m}_i\|_2}.
\end{equation}
Numerically, the normalized spectral coordinates are
\begingroup
\small
\begin{equation}
\begin{array}{c|ccc}
 & \widehat{m}_{i,1} & \widehat{m}_{i,2} & \widehat{m}_{i,3}\\
\hline
XX & 0.7136 & 0.6067 & 0.3503\\
YY & 0.7136 & -0.6067 & 0.3503\\
ZZ & 0.7136 & 0 & -0.7005\\
XI & 0.4046 & 0.7920 & 0.4573\\
YI & 0.4046 & -0.7920 & 0.4573\\
ZI & 0.4046 & 0 & -0.9145
\end{array}.
\end{equation}
\endgroup

The distance between two vertices in the spectral embedding is measured using the Euclidean distance
\begin{equation}
    d_{ij}
    =
    \left\|
    \widehat{\mathbf{m}}_i-\widehat{\mathbf{m}}_j
    \right\|_2 .
\end{equation}
The most relevant pairwise distances are
\begin{equation}
d
=
\vcenter{\hbox{\scriptsize
$
\left[
\begin{array}{c|cccccc}
 & XX & YY & ZZ & XI & YI & ZI \\
\hline
XX & 0 & 1.213 & 1.213 & 0.376 & 1.436 & 1.436 \\
YY & 1.213 & 0 & 1.213 & 1.436 & 0.376 & 1.436 \\
ZZ & 1.213 & 1.213 & 0 & 1.436 & 1.436 & 0.376 \\
XI & 0.376 & 1.436 & 1.436 & 0 & 1.584 & 1.584 \\
YI & 1.436 & 0.376 & 1.436 & 1.584 & 0 & 1.584 \\
ZI & 1.436 & 1.436 & 0.376 & 1.584 & 1.584 & 0
\end{array}
\right]
$
}}
.
\label{eq:spectral_distance_matrix}
\end{equation}
Thus, the nearest spectral-coordinate pair to $XX$ is $XI$, the nearest spectral-coordinate pair to $YY$ is $YI$, and the nearest spectral-coordinate pair to $ZZ$ is $ZI$. Therefore, applying $k$-means with $k=3$ to the row-normalized spectral embedding gives
\begin{align}
    \mathcal{C}_1 &= \{XX,XI\}, \nonumber\\
    \mathcal{C}_2 &= \{YY,YI\}, \nonumber\\
    \mathcal{C}_3 &= \{ZZ,ZI\}. \nonumber
\end{align}

Note that here, closeness does not mean a Hilbert-space distance between the Pauli operators as matrices. Instead, it means distance in the low-dimensional graph embedding obtained from the normalized Laplacian. Operators are close in this embedding when their vertices have similar commutation-affinity structure in the graph. Hence, the pairs $(XX,XI)$, $(YY,YI)$, and $(ZZ,ZI)$ are clustered together.


\section{Synthesizing Clifford circuits for joint measurements of commuting Pauli operators}
\label{sec:qc}

Here, we describe a procedure to synthesize quantum circuits comprising only elementary Clifford gates---the Hadamard gate $H$, the phase gate $S$, and the CNOT gate---for joint measurements of arbitrary sets of $N$-qubit mutually commuting Pauli operators. In this procedure, we utilize the binary symplectic representation (BSR) and the method of symplectic gaussian elimination (SGE)~\cite{aaronson-gottesman-PRA2004, crawford-quantum-2021, hasan-prxquantum-2025}. 

Given a set of mutually commuting Pauli operators $\mathcal{P} = \{P_1, P_2, \cdots, P_m\}$, the goal is to construct a Clifford circuit $U$ such that
\begin{equation}
    U P_j U^{\dagger} \rightarrow \{\pm I, \pm Z \}^{\otimes N} , \;\; \forall P_j\in\mathcal{P}.
\end{equation}
When we find such a circuit, all $P_j \in \mathcal{P}$ can be estimated using a single measurement setting, i.e., applying $U$ and measuring every qubit in the computational basis.


\subsection{Binary symplectic representation}

Using BSR, an $N$-qubit Pauli operator can be represented using two binary vectors, each of size $N$. The first vector stores the Pauli-$X$ component, while the second vector stores the Pauli-$Z$ component. For example, the single-qubit Pauli operators have the following BSRs:
\begin{equation}
    X \equiv (1 \mid 0); \quad  Z \equiv (0 \mid 1); \quad Y = iXZ \equiv (1 \mid 1).
\label{eq:BSR-example}
\end{equation}
Note that the BSR does not store overall phase or sign information [as can be seen for $Y$ in \eqref{eq:BSR-example}], but only specifies the type of Pauli operator. 

Similarly, an $N$-qubit Pauli operator can be represented using two binary vectors,
\begin{equation}
\mathbf x = (x_1,\ldots,x_N)\in \mathbb{F}_2^N,
\quad
\mathbf z = (z_1,\ldots,z_N)\in \mathbb{F}_2^N,
\end{equation}
such that an arbitrary $N$-qubit Pauli operator
\begin{equation}
P = i^k X^{\mathbf x}Z^{\mathbf z}, 
\end{equation}
where
\begin{equation}
X^{\mathbf x}
= X_1^{x_1}X_2^{x_2}\cdots X_N^{x_N},
\quad
Z^{\mathbf z}
= Z_1^{z_1}Z_2^{z_2}\cdots Z_N^{z_N},
\end{equation}
has the BSR
\begin{equation}
P \equiv (\mathbf x\mid \mathbf z)\in \mathbb{F}_2^{2N}.
\end{equation}
Consequently, a set of $m$ $N$-qubit Pauli operators can be represented using a binary matrix $M$, also referred to as a \textit{Pauli check matrix} (PCM) or \textit{stabilizer check matrix}, as
\begin{equation}
    M = [ \mathcal{X} \mid \mathcal{Z} ] \in \mathbb{F}^{m \times 2N}_2,
\end{equation}
where 
\begin{equation}
\mathcal{X} =
\begin{bmatrix}
- & \mathbf x_1 & - \\
- & \mathbf x_2 & - \\
& \vdots & \\
- & \mathbf x_m & -
\end{bmatrix},
\qquad
\mathcal{Z} =
\begin{bmatrix}
- & \mathbf z_1 & - \\
- & \mathbf z_2 & - \\
& \vdots & \\
- & \mathbf z_m & -
\end{bmatrix}.
\end{equation}

Using BSR, the multiplication of two Pauli operators corresponds to row operations in the binary matrix $M$:
\begin{equation}
P_iP_j \equiv (\mathbf{x}_i+ \mathbf{x}_j \mid   \mathbf{z}_i+ \mathbf{z}_j) \quad (\text{mod 2}) .
\end{equation}
Similarly, commutation and anti-commutation relations between $P_i$ and $P_j$ can be expressed through the symplectic inner product defined as
\begin{equation}
    \langle P_i,P_j \rangle_{\rm sp} =\mathbf{x}_i\cdot \mathbf{z}_j^T + \mathbf{z}_i\cdot \mathbf{x}_j^T \quad (\text{mod 2}),
\end{equation}
implying
\begin{align}
& \langle P_i,P_j \rangle_{\rm sp}=0
\quad\Longleftrightarrow\quad
P_i \text{ and } P_j \text{ commute} , \\ 
& \langle P_i,P_j \rangle_{\rm sp}=1
\quad\Longleftrightarrow\quad
P_i \text{ and } P_j \text{ anti-commute}.
\end{align}
Consequently, the mutual commutation relation of a set of Pauli matrices represented using the binary matrix $M = [ \mathcal{X} \mid \mathcal{Z} ] $ can be verified via
\begin{equation}
    \mathcal{X}\mathcal{Z}^{T} + \mathcal{Z}\mathcal{X}^{T} = 0 \quad (\text{mod 2}).
\end{equation}
In addition, the action of Clifford gates---$H$, $S$, and CNOT---on Pauli operators correspond to symplectic column operations,
\[ 
P_i \rightarrow U P_i U^{\dagger} \equiv (\mathbf{x}_i \mid \mathbf{z}_i) \rightarrow (\mathbf{x}'_i \mid \mathbf{z}'_i) .
\]


\subsection{Symplectic Gaussian elimination}

We now describe how SGE can be applied to a set of mutually commuting Pauli operators in the BSR to synthesize a Clifford circuit enabling joint measurement of the set. Given the set of mutually commuting Pauli operators $\mathcal{P} = \{P_1, P_2, \cdots, P_m\}$ and its corresponding binary check matrix $M$, the goal of SGE is to find a Clifford circuit $U$ such that
\begin{equation}
     M = [ \mathcal{X} \mid \mathcal{Z} ]  \xrightarrow[]{U} M' = [ 0 \mid \mathcal{Z}' ]
\end{equation}

Here, we only use three elementary Clifford gates---$H$, $S$, and CNOT---to synthesize this unitary operation. The binary update rules for these gates are
\begin{itemize}
\item Hadamard gate $H_q$:
\begin{equation}
x_q \leftrightarrow z_q .
\end{equation}

\item Phase gate $S_q$:
\begin{equation}
z_q \leftarrow z_q+x_q \quad \text{(mod 2)},
\qquad
x_q \leftarrow x_q .
\end{equation}

\item $\mathrm{CNOT}_{c\to t}$ gate, where $c$ denotes the control qubit and $t$ denotes the target qubit:
\begin{align}
& x_t \leftarrow x_t+x_c \quad \text{(mod 2)},
\qquad
x_c \leftarrow x_c, \\ 
& z_c \leftarrow z_c+z_t \quad \text{(mod 2)},
\qquad
z_t \leftarrow z_t . 
\end{align}
\end{itemize}

As an example, consider the following set of mutually commuting two-qubit Pauli operators:
\begin{equation}
    \mathcal{P} = \{XX, YY, ZZ \}.
\end{equation}
The corresponding Pauli check matrix $M$ is
\begin{equation}
    M_{\mathcal{P}} =
\begin{bmatrix}
1&1 \mid 0&0\\
1&1 \mid 1&1\\
0&0 \mid 1&1
\end{bmatrix}.
\end{equation}
After applying SGE we obtain
\begin{equation}
    M_{\mathcal{P}} \xrightarrow[\text{step 1}]{\mathrm{CNOT}_{1\to 2}}
\begin{bmatrix}
1&0 \mid 0&0\\
1&0 \mid 0&1\\
0&0 \mid 0&1
\end{bmatrix} \xrightarrow[\text{step 2}]{H_1} 
\begin{bmatrix}
0&0 \mid 1&0\\
0&0 \mid 1&1\\
0&0 \mid 0&1
\end{bmatrix} .
\end{equation}
Thus, the corresponding change-of-basis Clifford circuit is
\begin{equation}
    U = H_1 \text{CNOT}_{1\to 2},
\end{equation}
which maps
\begin{equation}
    XX \xrightarrow[]{U} ZI, \quad YY \xrightarrow[]{U} ZZ, \quad \text{and} \quad ZZ \xrightarrow[]{U} IZ,
\end{equation}
up to a global sign.
Consequently, a single experimental setting---applying $U= H_1 \text{CNOT}_{1\to 2}$ on the initial state $\rho$ followed by measurement in the computational basis---is enough to span and estimate all three Pauli operators $XX$, $YY$, and $ZZ$.

In Algorithm~\ref{alg:sge}, we provide a greedy algorithm for SGE. This greedy approach is constructive and guaranteed to produce a valid Clifford basis change for any mutually commuting Pauli set to synthesize the sought measurement circuit. However, the algorithm is not, in general, optimal in terms of the number of CNOT gates or the circuit depth. Several approaches have been developed to address this challenge by optimizing the SGE process to reduce both CNOT count and circuit depth~\cite{patel-quant-info-comp-2008, michael-pra-2023, jens-npj-2024}.

\begin{algorithm}[H]
\caption{Heuristic SGE for joint Pauli measurement}
\label{alg:sge}
\begin{algorithmic}[1]
\Require Commuting Pauli set $\mathcal P=\{P_1,\ldots,P_m\}$ on $N$ qubits
\Ensure Clifford $U$ such that 
\[U P_j U^{\dagger} \rightarrow \{\pm I, \pm Z \}^{\otimes N} , \;\; \forall P_j\in\mathcal{P} \]

\State Using SBR, encode $\mathcal P$ as a binary check matrix
\[
M=[\mathcal{X} \mid \mathcal{Z}]\in\mathbb F_2^{m\times 2N}.
\]
\State $U\gets I,\quad A\gets\{1,\ldots,N\}$.

\While{$\mathcal{X}$ has a nonzero entry in an active column}
    \State Choose a row $r$ and active pivot qubit $p\in A$ with $x_{r,p}=1$.
    \State Use local Cliffords on active qubits to map the support of row $r$ to $\mathcal{X}$'s.
    \State For each $q\neq p$ with $x_{r,q}=1$, apply $\mathrm{CNOT}_{p\to q}$.
    \State Apply $H_p$, so row $r$ becomes $\mathcal{Z}_p$.
    \State Update $M$ and append all gates to $U$.
    \State $A\gets A\setminus\{p\}$.
\EndWhile

\State \Return $U$
\end{algorithmic}
\end{algorithm}


\bibliography{ref}	


\end{document}